\documentclass[11pt]{article}

\usepackage[letterpaper,margin=1in]{geometry}
\usepackage{setspace}
\usepackage{placeins}
\usepackage{amsmath,amssymb,amsthm,mathtools}
\usepackage{graphicx}
\usepackage{booktabs}
\usepackage[expansion=false]{microtype}
\usepackage[dvipsnames]{xcolor}
\usepackage[authoryear,round]{natbib}
\usepackage[colorlinks=true,linkcolor=RoyalBlue,citecolor=RoyalBlue,urlcolor=RoyalBlue]{hyperref}

\theoremstyle{plain}
\newtheorem{theorem}{Theorem}[section]
\newtheorem{proposition}[theorem]{Proposition}
\newtheorem{lemma}[theorem]{Lemma}

\theoremstyle{definition}
\newtheorem{definition}[theorem]{Definition}
\newtheorem{assumption}[theorem]{Assumption}
\newtheorem{remark}[theorem]{Remark}

\newif\ifdnotes \dnotesfalse   % \dnotestrue shows the draft notes (see dnotes_1018.txt)

\newcommand{\OAproofs}{Online Appendix~A}
\newcommand{\OAsim}{Online Appendix~C}
\newcommand{\OAladder}{Online Appendix~D}
\newcommand{\OAtanaka}{Lemma~B.1 of the Online Appendix}
\newcommand{\OAremlimit}{Remark~B.2 of the Online Appendix}
\newcommand{\OAremarks}{Remarks~B.2 and B.3 of the Online Appendix}
\newcommand{\OAremsing}{Remark~B.3 of the Online Appendix}

\newcommand{\OAtabfull}{Table~E.1 of the Online Appendix}
\newcommand{\OAtabrobust}{Table~E.2 of the Online Appendix}
\newcommand{\caplag}{
The departure-hazard model is refitted using queue depth
measured $h$ seconds before the current instant. Panel (a) divides each estimate by its
true value, so that $1$ means the target is recovered; panel (b) shows the estimates
themselves, with the true values marked by short horizontal bars to the right of ``entry''. At $h=0$, depth is a contemporaneous
internal covariate; ``entry'' denotes depth measured at the
start of each visit. For level B, where persistence arises
through replenishment, contemporaneous adjustment largely
absorbs the effect: the estimated index is $1.006$, compared
with a true value of $1.882$. Estimates for levels C and D,
whose mechanisms affect departure intensity conditional on
queue state, remain comparatively stable across lags.
Within this simulation design, the lag profiles distinguish
the specified mechanisms.}
\newcommand{\capprofile}{Residual log-odds of an upward single-tick departure by position relative to the
nearest round dollar, 2025, with the stocks ordered by the relative size of the tick. Position
$0$ has the round dollar as the best bid and position $-1$ as the best ask (red); positions $-6$,
$-5$, $4$ and $5$ have a nickel as a quote (dark grey). Levels more than six ticks from a round
dollar are the reference. Controls are those of the direction equation. Bars are 95\%
intervals clustered on day. The Apple panel pools two half-samples (even and odd trading days)
by inverse-variance weighting.}
\newcommand{\capelapsed}{Round-dollar effects by time since entry.
Points show inverse departure hazard ratios,
$e^{-\hat{\gamma}_r(u)}$, from the piecewise-exponential
specification, with 95\% confidence intervals based on
day-clustered standard errors.
Circles and diamonds indicate visits with a round dollar
as the best bid and best ask, respectively.
Values below one indicate higher departure hazards than
at control levels, conditional on the model covariates;
the dashed line marks one.}

\newcommand{\R}{\mathbb{R}}
\newcommand{\Z}{\mathbb{Z}}
\newcommand{\E}{\mathbb{E}}
\newcommand{\PP}{\mathbb{P}}
\newcommand{\qv}[1]{[#1]}
\newcommand{\1}{\mathbf{1}}
\newcommand{\dd}{\mathrm{d}}

\title{Do Round Numbers Hold Prices?\\
Occupation Densities and Depth-Conditioned Hazards\\
in the Limit Order Book}

\author{Jiwon Jung\thanks{Corresponding author. Department of Applied Mathematics, University of Washington, 4182 W Stevens Way NE, Seattle, WA 98105, USA.
Tel.\ +1 206-543-5493. 
Email: jung320@uw.edu.}
\and
Kiseop Lee\thanks{Department of Statistics, Purdue University, 150 N.\
University Street, West Lafayette, IN 47907, USA. Email: kiseop@purdue.edu.}}

\date{\today}

\begin{document}

\maketitle
\thispagestyle{plain}
\setstretch{1.1}

% \maketitle
% \thispagestyle{plain}

% \newpage
% \doublespacing

\begin{abstract}
\noindent
Prices cluster at round numbers, a pattern often interpreted as evidence that these levels act as price barriers. This pattern also admits a mechanical explanation: more liquidity rests there, and prices take longer to leave deeper queues. Limit order book data allow us to distinguish these mechanisms through a tick-grid framework in which local time counts departures and level-wise volatility reflects the departure hazard. Using data for five Nasdaq stocks throughout 2025, we find that round levels attract more liquidity but do not exhibit longer depth-adjusted durations. Intel and SoFi show directional effects, with departures biased away from round levels.
\end{abstract}



% \newpage

%\tableofcontents

%==============================================================================
\section{Introduction}\label{sec:intro}
%==============================================================================

% \subsection{The question, the estimand, and the literature}

Prices cluster at round numbers. \citet{Osborne} documented concentration at whole-number prices, and \citet{Harris} found systematic clustering in both transaction prices and quotes. The phenomenon has been documented across a wide range of markets, with \citet{LallouacheAbergel} showing that limit orders also concentrate at round prices, resulting in greater displayed depth at these levels. Such concentrations of liquidity have been interpreted as price barriers, as accumulated depth may slow price movements through round levels.

The difficulty is that the evidence offered for a barrier often consists of the same marginal statistics used to document clustering. Excess last-digit frequencies, greater displayed depth, and longer residence times can all arise from a simple queueing mechanism. A preference for placing limit orders at round prices can generate deeper queues, which may take longer to deplete.
Prices may thus linger at round numbers because the queues are deeper, without any additional persistence attributable to roundness itself.  
The distinction is therefore \emph{conditional}: given the same amount of resting liquidity, does the price remain longer at a round level? Answering this question requires information about liquidity that price observations alone do not provide. Limit order book data provide this information by recording displayed depth alongside quoted prices. We therefore ask:
\begin{quote}
\emph{Conditional on visible liquidity, do prices remain longer at round price levels than at non-round levels?}
\end{quote}
 
We call an affirmative answer \emph{residual stickiness}, to distinguish it from the unadjusted persistence measured in the clustering literature. If longer residence at round numbers is entirely explained by differences in visible liquidity, no additional effect associated with roundness is needed to account for that persistence. If a residual remains, visible liquidity alone does not fully explain the observed persistence. Possible explanations include hidden and iceberg liquidity, option-related pinning, and behavioural reluctance to quote through a round level. These mechanisms are economically distinct, and distinguishing among them requires further evidence.

Answering the question requires specifying what it means for a price to ``linger'' at a level. We use two quantities: occupation time and departure count. Let $X_t$ be the best bid at the most recent time when the bid--ask spread was exactly one tick, so that $X_t$ takes values on a grid of tick size $\delta$. For a level $k$ and a window $[0,t]$, let $T_t^k$ denote the time spent at $k\delta$ and $N_t^k$ the number of departures from that level. In continuous-time finance, \emph{local time} describes how a diffusion's quadratic variation is distributed across price levels through an occupation identity. The level-wise volatility estimator of \citet{FlorensZmirou} relates this accumulated variation to the time spent at a level. On a tick grid, however, the price is constant between jumps: its continuous quadratic variation is zero, and the corresponding local time vanishes. We instead account for the quadratic variation carried by jumps. When price changes are restricted to one-tick moves, each departure contributes $\delta^2$ to quadratic variation, yielding an exact occupation identity with the discrete analogue $\ell_t^k=\delta N_t^k$. Dividing the variation accumulated at a level by the time spent there gives
\begin{equation}\label{eq}
\hat\sigma^2(k\delta) \;=\; \frac{\delta^2 N_t^k}{T_t^k},
\end{equation}
a tick-grid counterpart of the Florens-Zmirou estimator.

This representation also provides an interpretation in terms of queue dynamics. When price departures occur through depletion of the best bid or ask queue, $N_t^k/T_t^k$ measures their combined depletion rate, linking level-wise volatility to the time until queue depletion. This connection places the substantive question within a survival-analysis framework. Departures are events, the time to a departure is a duration, visible depth is a covariate, and round-number status identifies the price levels being compared. The estimand is the ratio of mean durations at matched entry depth, estimated using a proportional-hazards regression on a joint occupation measure over price level and queue state.

The paper connects several strands of literature. Studies of price clustering document regularities in last-digit frequencies and liquidity at round numbers. Our focus is whether these levels exhibit additional persistence after conditioning on visible liquidity. Models of order book dynamics provide a structural account of the relationship between queues and price changes: \citet{ContdeLarrard} develop a Markovian level-1 book and its diffusive limit, while \citet{HuangLehalleRosenbaum} introduce the queue-reactive model. The uncertainty-zones model of \citet{RobertRosenbaum} provides a related framework for price dynamics on a tick grid. We complement these approaches by linking queue state to a level-indexed occupation measure.
 
The statistical literature on local time provides estimators \citep{FlorensZmirou,JacodRates}, as well as convergence rates and limit theorems for skew and oscillating Brownian motion \citep{LejayMordeckiTorres,MazzonettoLejay}. Related work on sticky, skew and oscillating diffusions studies distinctive behaviour at particular levels \citep{EngelbertPeskir,OptimalStopping,SkewIndex}. Our contribution is to formulate local-time-based estimation directly on the tick grid and incorporate queue depth as a conditioning variable, allowing us to examine how much of a level's observed persistence is explained by the liquidity resting there.
Level-wise volatility estimates also enter diffusion-based approaches to bubble detection \citep{DelbaenShirakawa,JarrowKchiaProtter}. This connection provides an additional motivation for distinguishing variation associated with queue depth from residual effects associated with particular price levels.

% \subsection{Contributions, findings, and outline}

Queue-based simulations illustrate three specification issues. Flexible depth controls reduce bias caused by differences in depth distributions across levels. Mean-duration ratios differ from inverse hazard ratios when the duration baseline varies with elapsed time. Contemporaneous depth adjustment can also absorb persistence generated through replenishment, while the
lag profile helps distinguish the simulated mechanisms.
These issues connect the framework to established concerns about covariate specification and hazard interpretation in survival analysis \citep{KalbfleischPrentice}.

The paper contributes a framework that nests the conventional Florens-Zmirou estimator and the departure-share estimator of skew as its saturated null model (Proposition~\ref{prop:nest}). Simulations generated from queue dynamics rather than imposed hazards demonstrate recovery of planted effects and reveal the biases introduced by simpler specifications. An application to a year of order book data for five Nasdaq stocks shows that round levels attract more resting liquidity in all five stocks, but neither raw nor depth-adjusted durations indicate greater persistence. In Intel and SoFi, round levels instead exhibit a directional effect: prices depart sooner, with departures systematically biased away from the round number. Both stocks have comparatively large ticks relative to price. Within either stock, however, the directional effect does not strengthen as prices fall, suggesting that relative tick size alone is insufficient to explain the observed pattern.  

Section~\ref{sec:data} describes the data and the construction of the price series, visits, and covariates. Sections~\ref{sec:grid}--\ref{sec:depth} develop the framework, Section~\ref{sec:sim} evaluates its performance in simulations, and Section~\ref{sec:app} presents the empirical application. Section~\ref{sec:disc} concludes. Proofs, the scaling limit, two further simulation studies and further robustness tables are in the Online Appendix.

%==============================================================================
\section{Data and variable construction}\label{sec:data}
%==============================================================================

\subsection{Source and sample}
We use limit order book data from the Nasdaq TotalView-ITCH feed, obtained through Databento’s XNAS.ITCH dataset in the MBP-10 format. The data contain order book updates with nanosecond timestamps. Our analysis uses the first level of the order book, including the best bid and ask prices and their displayed sizes, to track changes in quoted prices and displayed depth.

The sample covers 250 trading days in 2025 for five Nasdaq-listed stocks: Apple, Cisco, Intel, SoFi Technologies, and American Airlines. Median prices range from \$12.69 to \$215.10. Relative tick sizes, measured as the one-cent tick divided by each stock's median price, range from 0.5 to 7.9 basis points. The spread equals one tick during 28.7\% of session time for Apple and 93.7--99.3\% for the other four stocks. The analysis covers the regular trading session, from 09:30 to 16:00 New York time, or to 13:00 on early-closing days.

\subsection{Constructing the price series}\label{sec:cleaning}

Identifying departures from price levels requires a price series that accounts for quote discreteness and spread fluctuations. A one-tick change in either best quote, with the other unchanged, shifts the midprice by half a tick. Changes between one- and two-tick spreads can therefore generate half-tick midprice movements even though quoted prices remain on the tick grid. Transaction prices can also alternate between the bid and ask without a change in either quote. We therefore construct the price process using observations at which the spread equals one tick.

Let $\delta$ denote the tick size, and let $b_t$ and $a_t$ denote the best bid and ask prices in ticks at time $t$. We call $t$ a \emph{one-tick moment} if $a_t-b_t=1$. Within each uninterrupted observation period, define
\[
X_t=\delta b_{\sigma_t},
\qquad
\sigma_t=\sup\{s\le t:a_s-b_s=1\},
\]
where $s$ ranges over recorded book update times in that period. The process is initialised at the first one-tick moment. It is piecewise constant, takes values in $\delta\mathbb{Z}$, and changes whenever the best bid at a subsequent one-tick moment differs from its previously recorded value. If the spread widens and then returns to one tick with the same best bid, $X_t$ remains unchanged.

The uncertainty-zones model of \citet{RobertRosenbaum} links transaction price changes on the tick grid to movements in a latent efficient price. Here, we construct a price process directly from observed quotes using a deterministic rule, without specifying a latent price process.

Observation is interrupted by a trading halt, a spread exceeding three ticks, or a spread remaining above one tick for more than sixty seconds. After an interruption, observation resumes at the next one-tick moment. We report the fraction of session time excluded and vary the spread cutoffs in robustness checks.

\subsection{Visits and covariates}\label{sec:visits}

A \emph{visit} is a maximal observed interval during which $X_t$ remains at a fixed price level. We index the retained visits across stocks and trading days by $i=1,\dots,n$. Visit $i$ begins at time $t_i$, occurs at price $k(i)\delta$, and has observed duration $\tau_i$. At each one-tick moment within the visit, the best bid and ask are $k(i)\delta$ and $(k(i)+1)\delta$. Thus, the visit corresponds to a pair of adjacent quotes and is indexed by its best bid.

An observed departure occurs when $X_t$ changes within an uninterrupted observation period. We set $D_i=1$ when a departure is observed and $D_i=0$ when observation is interrupted before a departure by the session close, a trading halt, or a spread cutoff. In the latter case, the visit is \emph{right-censored}, and $\tau_i$ records its duration up to the interruption. For a spread exclusion, censoring occurs when the cutoff is crossed. Visits with unobserved entry times, including the first visit of each session and the first after an excluded period, are omitted. 

For a one-tick departure, $J_i\in\{+,-\}$ indicates an upward or downward move. Visits ending in multi-tick moves are retained as completed visits in the duration analysis but excluded from the direction analysis. We report the proportions of visits that are censored or end in multi-tick moves.
 
Let $u$ denote elapsed time since entry into a visit. For $0\le u<\tau_i$, the covariate path $Z_i(u)$ records the displayed depths at the best bid and ask in shares, their imbalance, the prevailing spread, and recent realised volatility at time $t_i+u$. These variables are updated at each book update. Their entry values are denoted by $Z_i=Z_i(0)$, and the corresponding bid and ask depths are referred to as \emph{entry depths}.

\subsection{Level types}\label{sec:levels} 

We classify price levels by their relation to round prices and pool visits with the same classification across absolute price levels. We consider multiples of \$1.00, \$0.50, \$0.25, and \$0.10, assigning each quote to the coarsest price increment it satisfies so that the categories are mutually exclusive. These categories allow us to examine whether persistence varies with the degree of roundness. We also consider placebo levels at fixed offsets from whole dollars, such as \$0.03 above each dollar, to assess whether similar patterns arise at non-round prices.

For each round-price category, we distinguish visits in which the round price appears as the best bid from those in which it appears as the best ask at the one-tick moments defining the visit. This distinction allows the effects associated with round prices on the bid and ask sides to be estimated separately. 

For a given category and quote side, the indicator $R_k$ records whether level $k$ belongs to that type. The reference group consists of levels assigned to none of the specified types. We also include distance in ticks to the nearest whole-dollar price as a categorical covariate to account for differences associated with proximity to round-dollar levels.

%==============================================================================

\section{Occupation on the tick grid}\label{sec:grid}
%==============================================================================

\subsection{Departure counts as local time}

Fix a tick size $\delta > 0$ and let $\mathbb{S}_\delta = \delta\Z$. Let $X=(X_t)_{t\ge0}$ denote the tick-grid price process constructed in Section~\ref{sec:cleaning}. We regard $X$ as a c\`adl\`ag, $\mathbb{S}_\delta$-valued process on a filtered probability space $(\Omega,\mathcal{F},(\mathcal{F}_t),\PP)$. 

\begin{assumption}[Grid dynamics]\label{ass:grid}
$X$ is a finite-activity pure-jump semimartingale with
$\Delta X_s \in \{-\delta, 0, +\delta\}$ for all $s$, and $X_0 \in \mathbb{S}_\delta$.
\end{assumption}

The nearest-neighbour restriction yields the exact identities developed below. In the empirical analysis, visits ending in multi-tick departures are retained for duration analysis but excluded from directional statistics, as described in Section~\ref{sec:visits}.  
Write
\[
  N^k_t \;=\; \#\{ s \le t:\ X_{s-} = k\delta,\ \Delta X_s \neq 0\},
  \qquad
  T^k_t \;=\; \int_0^t \1\{X_s = k\delta\}\,\dd s
\]
for the \emph{departure count} and the \emph{occupation time} at level $k$, and
split the former by direction, $N^k = N^{k,+} + N^{k,-}$.

We first establish that the semimartingale local time vanishes for the tick-grid process. 

\begin{lemma}\label{lem:vanish}
Under Assumption~\ref{ass:grid}, $\qv{X}^c \equiv 0$ and hence the semimartingale
local time $L^a_t \equiv 0$ for every $a \in \R$, $\PP$-a.s.
\end{lemma}

\begin{proof}
A c\`adl\`ag path with values in the discrete set $\mathbb{S}_\delta$ has
finitely many jumps on compacts, so $X$ is piecewise constant and of finite
variation; its continuous martingale part is therefore null and
$\qv{X}_t = \sum_{s\le t}(\Delta X_s)^2$ is a pure-jump process, i.e.\
$\qv{X}^c \equiv 0$. Semimartingale local time satisfies the occupation-density
formula $\int_0^t f(X_{s-})\,\dd\qv{X}^c_s = \int_\R f(a)L^a_t\,\dd a$ for bounded
measurable $f$; the left side is identically zero, so $L^{\cdot}_t = 0$ a.e.\ in
$a$, and right-continuity in $a$ gives the claim for every $a$.
\end{proof}

The calendar-time occupation measure is likewise discrete:
\[
  \sum_k T^k_t\,\varepsilon_{k\delta}.
\]
It is supported on the tick grid and therefore has no density with respect to Lebesgue measure. A grid-based formulation must instead assign occupation mass to individual price levels.

We define discrete local time by scaling the departure count by the tick size. This construction is related to classical approximations of local time based on visits and crossings, including L\'evy's downcrossing theorem \citep[Ch.~VI]{RevuzYor}, invariance principles for random-walk local time \citep{Knight1963,Borodin1981}, 
and crossing-count estimators of local time \citep{MazzonettoLejay}. Under Assumption~\ref{ass:grid}, the resulting quantity satisfies an exact occupation identity with respect to $\qv{X}$, as shown in Lemma~\ref{lem:occ}.

\begin{definition}[Discrete local time]\label{def:ell}
$\displaystyle \ell^k_t \;:=\; \delta\,N^k_t$.
\end{definition}

\begin{lemma}[Occupation identity against $\qv{X}$]\label{lem:occ}
Under Assumption~\ref{ass:grid}, for every bounded measurable $f$ and every $t$,
\begin{equation}\label{eq:occ}
  \int_0^t f(X_{s-})\,\dd\qv{X}_s \;=\; \sum_{k \in \Z} f(k\delta)\,\ell^k_t\,\delta ,
  \qquad\text{and in particular}\qquad
  \qv{X}_t \;=\; \delta \sum_{k\in\Z} \ell^k_t .
\end{equation}
\end{lemma}

\begin{proof}
With the convention $\qv{X}_0 = 0$ and $\qv{X}^c\equiv 0$ (Lemma~\ref{lem:vanish}),
$$\int_0^t f(X_{s-})\dd\qv{X}_s = \sum_{s\le t} f(X_{s-})(\Delta X_s)^2
 = \delta^2 \sum_k f(k\delta) N^k_t = \sum_k f(k\delta)(\delta N^k_t)\delta.$$
Take $f \equiv 1$ for the second identity.
\end{proof}

An alternative construction counts crossings of half-tick levels. Let $C^a_t$ denote the number of crossings of $a=(k+\tfrac12)\delta$. These counts enter the grid version of the Tanaka--Meyer formula (\OAtanaka) and satisfy
\[
C^{(k+1/2)\delta}_t=N^{k,+}_t+N^{k+1,-}_t.
\]
Thus, half-tick crossing counts combine departures from adjacent sites, whereas $\ell^k_t$ records departures from a single site. Their asymptotic relationship depends on local directional symmetry and spatial regularity, conditions that need not hold at a skew level (\OAremsing). We use the site-indexed quantity $\ell^k_t$ because it satisfies \eqref{eq:occ} exactly at each level.

\subsection{Occupation measures under three clocks}\label{sec:clocks}

A \emph{clock} is a non-decreasing process $A_t$ by which activity is measured,
and each clock carries a level-indexed occupation measure
$O^k_t(A) = \int_0^t \1\{X_{s-}=k\delta\}\,\dd A_s$, the amount of the clock
accumulated while the price is at level $k$. The three natural clocks are
calendar time $A_t=t$, cumulative traded volume $A_t=V_t$, and quadratic variation
$A_t=\qv{X}_t$, and they give
\[
\begin{array}{llll}
  \text{calendar:} & T^k_t &=& \int_0^t \1\{X_s=k\delta\}\,\dd s \;=\; \text{seconds spent with the price at } k\delta,\\[2pt]
  \text{volume:}   & V^k_t &=& \text{shares traded with the price at } k\delta,\\[2pt]
  \text{quadratic variation:} & \delta\,\ell^k_t &=& \delta^2 N^k_t .
\end{array}
\]

For calendar time, $X_{s-}$ and $X_s$ differ only at the countably many jump times,
so $O^k_t(t)$ is the $T^k_t$ above. For the other two clocks, which put mass at
jump times, $X_{s-}$ attributes the mass of a jump or a trade to the level it
leaves. The last line is $O^k_t(\qv{X})$: each departure from level $k$ adds $\delta^2$ to
$\qv{X}$, and $\ell^k_t = \delta N^k_t$ is this quantity per unit of the grid
spacing $\delta$, which is the sense in which it is a density
(Lemma~\ref{lem:occ}). 

The calendar-time and volume occupation measures describe how time and trading activity are distributed across price levels. The quadratic-variation occupation measure additionally yields a level-wise variance rate. For $T^k_t>0$,
\begin{equation}\label{eq:rn}
  \frac{\delta\,\ell^k_t}{T^k_t} \;=\; \frac{\delta^2 N^k_t}{T^k_t} \;=\; \hat\sigma^2(k\delta).
\end{equation}
This ratio is the level-wise volatility estimator used below.

\begin{remark}[Queue-depletion interpretation]\label{rem:hazard}
The estimator $\hat\sigma^2(k\delta)$ equals $\delta^2$ times the observed departure rate $N^k_t/T^k_t$. When departures correspond to depletion of the best bid or ask queue, this rate measures their combined depletion frequency per unit of occupation time. The representation links level-wise volatility to queue dynamics and motivates conditioning departure rates on visible depth in Section~\ref{sec:depth}. 
\end{remark}

%==============================================================================

\subsection{Level-wise statistics}\label{sec:est}
%==============================================================================

Fix an observation window $[0,t]$; quantities may also be aggregated across trading days. We consider three features of price dynamics at a level: the price may stay there unusually long (a \emph{sticky} level), leave it asymmetrically (a \emph{skew} level), or change its volatility on crossing it (an \emph{oscillating} level). Each has a continuous-time prototype and a statistic on the grid.

A sticky point is a level at which a diffusion can spend positive calendar time without being absorbed. For sticky Brownian motion, this behaviour is described by an occupation-time relation of the form
\[
\int_0^t \1\{X_s=x_0\}\,\dd s=\theta L^{x_0}_t,
\]
where $\theta>0$ measures stickiness under the chosen normalisation of local time \citep{EngelbertPeskir}. This relation motivates a grid-based statistic formed from occupation time $T^k_t$ and discrete local time $\ell^k_t$.

\begin{definition}[Stickiness statistic]\label{def:theta}
$\displaystyle \hat\theta_k \;:=\; \frac{T^k_t}{\ell^k_t} \;=\; \frac{T^k_t}{\delta N^k_t}$.
\end{definition}

By \eqref{eq:rn}, $\hat\theta_k=\delta/\hat\sigma^2(k\delta)$. This relation provides a benchmark for interpreting the statistic as the grid is refined. Under the sampling and convergence conditions discussed in \OAremarks, a regular level with strictly positive limiting variance has $\hat\theta_k$ of order $\delta$, whereas a sticky level can have a positive limiting occupation-to-local-time ratio. The distinction therefore concerns whether this ratio vanishes as $\delta\to0$ or remains positive.

At a fixed tick size, we assess persistence relative to a set of control levels $\mathcal{K}$, such as neighbouring levels or non-round levels for the same stock and trading day. Define
\[
\bar\theta
=\frac{\sum_{j\in\mathcal{K}}T^j_t}
{\delta\sum_{j\in\mathcal{K}}N^j_t}.
\]
The ratio $\hat\theta_k/\bar\theta$ measures unadjusted persistence at level $k$ relative to these controls. Section~\ref{sec:depth} extends this comparison by conditioning on visible liquidity.

% In data, $\delta$ is fixed and the absolute statement is not available; the
% operational version is relative. Let $\mathcal{K}$ be a set of control levels (a
% neighbourhood of $k$, or the pooled non-special levels of the same name and day)
% and set $\bar\theta = \sum_{j\in\mathcal{K}} T^j_t \,/\, \delta\sum_{j\in\mathcal{K}} N^j_t$. The raw
% statistic is the ratio $\hat\theta_k/\bar\theta$. Section~\ref{sec:depth} refines
% it.

For skew Brownian motion with parameter $\alpha\in(0,1)$, excursions from the skew point are directed upward with probability $\alpha$ \citep{Lejay}. The corresponding grid statistic is the proportion of departures from a level that are upward.
% ``support'' is $\alpha > 1/2$ and ``resistance'' $\alpha<1/2$.

\begin{definition}[Skew statistic]\label{def:alpha}
$\displaystyle \hat\alpha_k \;:=\; N^{k,+}_t / N^k_t$, with
$\widehat{\mathrm{se}}(\hat\alpha_k) = \sqrt{\hat\alpha_k(1-\hat\alpha_k)/N^k_t}$.
\end{definition}

Under an independent Bernoulli model for departure directions, $\hat\alpha_k$ is the maximum-likelihood estimator of the upward-departure probability, and the displayed standard error is the corresponding binomial standard error. Inference for skew diffusion models under high-frequency
observation is studied by \citet{LejayMordeckiTorres}
and \citet{MazzonettoLejay}.
% The identification concern is drift: a drifting price also departs levels
% asymmetrically. On the grid, the two are separated by their order in $\delta$ and
% by their profile in $k$.

Directional asymmetry can also arise from drift. The following proposition distinguishes the local departure asymmetry induced by smooth drift from that associated with a skew point through its magnitude as $\delta\to0$ and its variation across neighbouring levels.

\begin{proposition}[Drift does not mimic skew]\label{prop:drift}
Let $X^0$ solve $\dd X^0 = b(X^0)\dd t + \sigma(X^0)\dd W$ with $X^0_0\in\delta\Z$,
where $\sigma^2$ is bounded away from zero and $\kappa := b/\sigma^2$ is Lipschitz on
a neighbourhood of $k\delta$. Observe $X^0$ on the grid through its skeleton:
$\rho_0 = 0$, $\rho_{n+1} = \inf\{t>\rho_n:\ |X^0_t - X^0_{\rho_n}| = \delta\}$,
so that departures from site $k$ are the steps of the chain
$(X^0_{\rho_n})_{n\ge0}$ out of $k\delta$. Then the directions of successive
departures from $k$ are i.i.d.\ Bernoulli$(p^+_k)$, so that
$\hat\alpha_k \to p^+_k$ almost surely as $N^k_t\to\infty$, and
\[
  p^+_k - \tfrac12 \;=\; \frac{b(k\delta)}{2\sigma^2(k\delta)}\,\delta \;+\; O(\delta^2),
\]
which vanishes with the grid and varies smoothly in $k$. If instead $k\delta$ is a
skew point of parameter $\alpha$ (and $X^0$ is a skew Brownian motion near it),
then $p^+_k = \alpha$ exactly, while $p^+_{k\pm1} = \tfrac12$: an $O(1)$ offset
localised at a single site.
\end{proposition}

The proof, which rests on the scale function of the diffusion, is in \OAproofs.

The proposition motivates assessing whether $\hat\alpha_k$ deviates from a smooth local trend across price levels, rather than testing only whether it differs from $1/2$. A local-polynomial fit provides one way to estimate this trend. The simulation in \OAsim{} illustrates the distinction between smooth drift-induced asymmetry and a localised skew effect.

An oscillating level separates regions with different volatility coefficients, as in oscillating Brownian motion \citep{MazzonettoLejay}. A corresponding grid statistic compares the variance rates in \eqref{eq:rn}, pooled separately above and below the threshold. We do not pursue this case here.

%==============================================================================

\section{Depth-conditioned occupation and the hazard model}\label{sec:depth}
%==============================================================================

\subsection{The joint occupation measure}

Let $Q_t$ denote the observable queue state at the prevailing level: the depths
resting at the best bid and at the best ask. 
The binned decomposition below uses total depth as a scalar summary, whereas the regression in Section~\ref{sec:ph} includes bid and ask depths separately.
Consider the joint process $Y = (X, Q)$ and define, for a depth bin $q$,
\[
  N^{k,q}_t = \#\{s \le t:\ X_{s-} = k\delta,\ Q_{s-} \in q,\ \Delta X_s \neq 0\},
  \qquad
  T^{k,q}_t = \int_0^t \1\{X_s = k\delta,\ Q_s \in q\}\,\dd s,
\]
and $\ell^{k,q}_t = \delta N^{k,q}_t$. 
The occupation identity in Lemma~\ref{lem:occ} extends to functions of both coordinates of $Y$, so $\ell^{k,q}$ is the occupation density
of the joint process; the price-marginal local time is the depth-aggregate
$\ell^k = \sum_q \ell^{k,q}$. The conditional statistics are
\[
  \hat\theta_k(q) = \frac{T^{k,q}_t}{\delta N^{k,q}_t},
  \qquad
  \hat\sigma^2(k\delta \mid q) = \frac{\delta^2 N^{k,q}_t}{T^{k,q}_t},
  \qquad
  \hat\alpha_k(q) = \frac{N^{k,q,+}_t}{N^{k,q}_t}.
\]
By Remark~\ref{rem:hazard}, $\hat\sigma^2(k\delta\mid q)$ is the depletion hazard
of a queue of size $q$ at price level $k\delta$: the queue-reactive object
\citep{HuangLehalleRosenbaum} indexed additionally by price level.

\subsection{Composition versus residual}

Since $T^k = \sum_q T^{k,q} = \delta\sum_q N^{k,q}\hat\theta_k(q)$, the raw
statistic decomposes exactly as a weighted average of conditional ones:
\begin{equation}\label{eq:decomp}
  \hat\theta_k \;=\; \sum_q w_k(q)\,\hat\theta_k(q),
  \qquad w_k(q) := \frac{N^{k,q}_t}{N^k_t}.
\end{equation}
Let $\bar\theta(q)$ be the pooled conditional stickiness at depth $q$ over the
control levels $\mathcal{K}$, 
and define the \emph{counterfactual} stickiness of level $k$ by combining its depth composition with the conditional stickiness of the control levels:
\[
  \hat\theta^{\,\mathrm{cf}}_k \;:=\; \sum_q w_k(q)\,\bar\theta(q).
\]
This gives a multiplicative decomposition analogous to a Blinder--Oaxaca decomposition or a stratified comparison, 
\begin{equation}\label{eq:BO}
  \underbrace{\frac{\hat\theta_k}{\bar\theta}}_{\text{raw}}
  \;=\;
  \underbrace{\frac{\hat\theta^{\,\mathrm{cf}}_k}{\bar\theta}}_{\text{composition}}
  \;\times\;
  \underbrace{\frac{\hat\theta_k}{\hat\theta^{\,\mathrm{cf}}_k}}_{\text{residual}} .
\end{equation}

The factors answer different questions, and the second captures persistence associated with the price level after conditioning on visible depth. A composition effect above one means the level is sticky because thick
queues rest there: the price does linger, but visible liquidity fully explains it, and a model that conditions on depth already accounts for this persistence. A residual effect above one means that, at
the same visible depth, this level's queues deplete more slowly than queues elsewhere;
the candidate explanations are hidden or iceberg liquidity, replenishment at round
numbers, option-related pinning, and behavioural reluctance to quote through the level.
Testing whether the residual differs from one requires observations of $Q$, which the price path alone does not provide. Order book data are therefore necessary for this conditional comparison.

The decomposition \eqref{eq:BO} provides a conceptual framework, but its direct estimation has several limitations.
It requires binning a continuous covariate, it has no natural
standard error, it cannot absorb intraday seasonality, and --- as
Section~\ref{sec:sim} shows --- it can be biased in the setting it is intended to address. The remainder of this section introduces a hazard regression that targets the same estimand, makes the underlying assumption of \eqref{eq:BO} explicit, and provides a means of testing it.

\subsection{The proportional-hazards specification}\label{sec:ph}

\subsubsection{Specification}

Visits, with their durations $\tau_i$, censoring indicators $D_i$, departure
directions $J_i$ and covariate paths $Z_i(\cdot)$, are as defined in
Section~\ref{sec:data}. Write $u$ for time elapsed since entry to the level (the
\emph{duration clock}, distinct from calendar time), and let $\lambda^{+}_i(u)$ and $\lambda^{-}_i(u)$ denote the cause-specific intensities of upward and downward departures. We use the equivalent parametrisation 
\begin{equation}\label{eq:param}
  \lambda_i(u) \;=\; \lambda^{+}_i(u) + \lambda^{-}_i(u),
  \qquad
  \pi_i(u) \;=\; \frac{\lambda^{+}_i(u)}{\lambda_i(u)} ,
\end{equation}
under which the likelihood of the visit data factorises,
\begin{equation}\label{eq:factor}
  \mathcal{L}
  \;=\;
  \underbrace{\prod_i \lambda_i(\tau_i)^{D_i}\exp\Big\{-\!\int_0^{\tau_i}\!\lambda_i(u)\,\dd u\Big\}}_{\text{duration}}
  \;\times\;
  \underbrace{\prod_{i:\,D_i=1} \pi_i(\tau_i)^{\1\{J_i=+\}}\big(1-\pi_i(\tau_i)\big)^{\1\{J_i=-\}}}_{\text{direction}} ,
\end{equation}
so that the two components are variation independent and may be estimated
separately with no loss of efficiency.

The parametrisation in \eqref{eq:param} corresponds directly to the two phenomena described in Section~\ref{sec:est}:
$\hat\sigma^2(k\delta)=\delta^2\hat\lambda_k$ and $\hat\theta_k=1/(\delta\hat\lambda_k)$ 
depend on the all-cause hazard, and $\hat\alpha_k=\hat\pi_k$ depends on the direction probability. Separate proportional-hazards models for $\lambda^{+}$ and $\lambda^{-}$ would couple these two components, since a sum of proportional hazards with different covariate
effects is not itself proportional.

The model is then, for visit $i$ in stratum $s(i)$, with time-varying covariates $Z_i(u)$ and
level-type design vector $R_{k(i)}$,
\begin{align}
  \lambda_i(u) &= \lambda_0\big(u \mid s(i)\big)\,
                  \exp\big\{\beta'Z_i(u) + \gamma'R_{k(i)}\big\},
  \label{eq:ph-haz}\\[2pt]
  \operatorname{logit}\pi_i(u) &= \eta_0\big(s(i)\big) + \psi'Z_i(u) + \phi'R_{k(i)}
                  + \varrho\,m(i),
  \label{eq:ph-dir}
\end{align}
where $m(i)$ is the signed distance in ticks from a reference price, included to account for drift when estimating skew (Proposition~\ref{prop:drift}).

\paragraph{Strata and the duration clock.} $s(i)=(\text{stock},\text{day},\text{time-of-day
bucket})$, with $\lambda_0(\cdot\mid s)$ left unspecified within stratum, so that the
intraday U-shape, the day's volatility level and stock-specific differences in departure rates are absorbed
non-parametrically; this provides additional flexibility relative to \eqref{eq:BO}. $\lambda_0$ is a
baseline in $u$, the time since arrival at the level. Section~\ref{sec:est} implicitly
assumes it is flat within a visit; \eqref{eq:ph-haz} does not, and a decreasing baseline ---
a departure rate that declines with elapsed residence time --- is the
signature of queue replenishment and a form of stickiness invisible to $\hat\theta_k$.
 
\paragraph{Covariates.} The covariate vector $Z$ may include log depth resting at the best bid and at the best ask (the two queues whose exhaustion ends a visit), queue imbalance, prevailing spread, and short-horizon realised volatility; Section~\ref{sec:app-spec} lists the covariates used in the application.
The level-type design vector $R_k$ includes indicators for round-dollar, half-dollar, quarter-dollar, and dime levels. Additional controls include distance in ticks to the nearest round dollar as a categorical variable and indicators for quotes within one tick of the previous closing price and the day's opening price. A smooth price-distance term accounts for residual spatial variation in departure dynamics.

\paragraph{Estimands.} The residual stickiness index of level type $r$ is the
ratio of mean durations at matched entry depth,
\begin{equation}\label{eq:estimands}
  S_r \;=\;
  \frac{\E\big[\,\tau \,\big|\, \text{level type } r,\ Z(0) = z\,\big]}
       {\E\big[\,\tau \,\big|\, \text{control},\ Z(0) = z\,\big]}
  \quad\text{averaged over } z,
  \qquad
  \underbrace{\phi_r}_{\text{residual skew, in log-odds}} .
\end{equation}
$S_r$ is what \eqref{eq:BO} estimates and what $\hat\theta_k$ measures, since $\hat\theta_k$ is
a mean duration per departure.

\begin{remark}[Hazard contrasts and mean-duration ratios]
\label{rem:estimand}
The inverse hazard ratio $e^{-\gamma_r}$ generally differs
from the mean-duration ratio $S_r$. Under the entry-depth
specification with a constant baseline within each stratum,
the two coincide. With a duration-dependent baseline,
$e^{\gamma_r}$ compares departure hazards conditional on
survival to matched elapsed time, whereas $S_r$ compares
mean durations at matched entry depth. In \OAladder,
for example, $e^{-\gamma_B}=2.84$ while $S_B=1.88$.
Estimating $S_r$ therefore requires integrating the fitted
survival functions and applying the averaging rule in its
definition, rather than exponentiating the coefficient.
\end{remark}

\subsubsection{A nesting hierarchy}

\begin{proposition}[Earlier estimators as special cases]\label{prop:nest}
Under Assumption~\ref{ass:grid}:
\begin{enumerate}
\item[(i)] Let $R$ be saturated in $k$, with no covariates or strata and a constant baseline in \eqref{eq:ph-haz}, and in \eqref{eq:ph-dir} drop $\eta_0(\cdot)$ and
  $\varrho\,m(i)$ as well. Then the maximum-likelihood estimator of the level-$k$
  hazard is $\hat\lambda_k = N^k_t / T^k_t$, so that
  $\delta^2\hat\lambda_k = \hat\sigma^2(k\delta)$ and
  $1/(\delta\hat\lambda_k) = \hat\theta_k$, and the maximum-likelihood estimator of
  the direction probability is $\hat\pi_k = \hat\alpha_k$. \emph{The Florens-Zmirou
  estimator and the statistics of Definitions~\ref{def:theta} and \ref{def:alpha}
  are the saturated null model of \eqref{eq:ph-haz}--\eqref{eq:ph-dir}.}
\item[(ii)] Let the depth bins $q$ be finitely many, with $Q$ measured
  contemporaneously as in the definition of $N^{k,q}, T^{k,q}$. Suppose the
  departure intensity at $(k,q)$ is constant --- $\lambda_0(q)$ at every control
  level and $\lambda_0(q)e^{\gamma_r}$ at every level of type $r$ --- i.e.\
  \eqref{eq:ph-haz} with baseline stratified by $q$, \emph{constant in the duration
  clock}, and one coefficient per level type. Then, as $T^{k,q}_t\to\infty$ for
  every $k$ and $q$, the residual factor of \eqref{eq:BO} converges to
  $e^{-\gamma_r}$ almost surely, whatever the weights $w_k(\cdot)$. If instead the
  coefficient is bin-specific, $\gamma_r(q)$, the residual factor converges to a
  weighted average of $e^{-\gamma_r(q)}$ whose weights depend on the level's own
  depth composition, and is therefore not comparable across levels.
\end{enumerate}
\end{proposition}

The proof is in \OAproofs.

The duration-clock restriction in (ii) is substantive: with a non-constant baseline a
ratio of mean durations is no longer $e^{-\gamma_r}$ (Remark~\ref{rem:estimand}). The use of contemporaneous depth is also substantive: the decomposition \eqref{eq:BO} is the lag-zero version
of the internal-covariate problem of Section~\ref{sec:endog}. The proposition shows
that Sections~\ref{sec:est} and \ref{sec:depth} correspond to special cases of the regression, the first with no controls and the second under an untested
proportionality assumption in depth, which part (ii) makes explicit and the diagnostics below
make testable. Violations of this assumption can be economically informative. Persistence concentrated at levels with thin \emph{visible} queues is consistent with hidden or iceberg liquidity.

\subsubsection{Estimation}\label{sec:ph-est}

For computational efficiency, we use a \emph{piecewise-exponential}
representation of the duration model.
We partition the duration axis at
$0 = u_0 < u_1 < \dots < u_J$ (duration quantiles), split each visit into the
cells it spans, and record for each cell an exposure and an event indicator. If
$\lambda_0(\cdot\mid s)$ is constant on each cell, the duration factor of
\eqref{eq:factor} is exactly the likelihood of a Poisson regression with
log-exposure offset,
\begin{equation}\label{eq:poisson}
  y_{ij} \sim \mathrm{Poisson}\big(\exp\{\log \Delta_{ij} + \nu_{js(i)}
  + \beta'Z_i + \gamma'R_{k(i)}\}\big),
\end{equation}
where $\Delta_{ij}$ is the time visit $i$ spends in cell $j$ and $\nu_{js}$ is the
cell-by-stratum baseline. This GLM formulation accommodates hundreds of millions of observations and is exact for a piecewise-constant baseline. Equation~\eqref{eq:ph-dir} is estimated using conditional logistic regression with one observation per visit.

\subsubsection{The endogenous-covariate problem}\label{sec:endog}

Queue depth at the prevailing level is an \emph{internal} covariate: it is generated
by the same order flow that produces the departure and is driven toward zero at the
moment of departure, so a model that conditions on $Q(u)$ as $u$ approaches $\tau_i$
conditions on the event itself. We consider four approaches.

\begin{enumerate}
\item[(a)] \textbf{Entry depth} (main specification). Hold $Z_i$ fixed at its value upon entry to the level, making it external to the visit. The estimand is then well
  posed: \emph{conditional on the same visible liquidity at entry, does the price take longer to leave the level of interest than a control level?}
\item[(b)] \textbf{Lagged depth} (\emph{diagnostic only}). $Z_i(u)=Q(u-h)$ for a fixed
  lag $h$. Varying $h$ does not provide a conventional robustness check for specification (a): at $h=0$ the model
  conditions on the state that produces the event, as $h$ grows it approaches the entry-depth fit (a), and in
  between it is consistent for neither (Section~\ref{sec:sim-qr}). The lag profile is used to distinguish between mechanisms: a level whose estimated effect persists as $h\to0$ acts on the departure intensity given the queue state, and one whose estimated effect approaches zero acts on how the queue evolves, that is, on replenishment
  (Figure~\ref{fig:lag}).
\item[(c)] \textbf{Entry depth with a duration-varying coefficient}
  (recommended for the diagnostic model). Keep $Z_i=Z_i(0)$ but let its coefficient
  depend on the duration clock,
  \begin{equation}\label{eq:optd}
    \lambda_i(u) \;=\; \lambda_0(u\mid s(i))\,
      \exp\big\{\beta(u)'Z_i(0) + \gamma'R_{k(i)}\big\}.
  \end{equation}
  This specification introduces no time-varying internal covariate, since $u$ is the baseline time scale. It also addresses a limitation of specification (a): the hazard of a hitting time given an \emph{external} initial state is
  a deterministic function of that state and elapsed time, so a constant depth
  coefficient is misspecified by construction, and what it fails to explain is
  absorbed by the level indicator. Specification (a) is the constant-coefficient special case. The simulation in \OAladder{} shows that (c) removes seven-eighths of the artefact while
  leaving genuine duration dependence intact.
\item[(d)] \textbf{Joint model.} Treat $(X,Q)$ as a multivariate counting process and
  the departure as the hitting time of $Q$ at zero, so that stickiness is a property
  of the replenishment intensity rather than a hazard ratio. This approach offers a more direct mechanistic interpretation but is more challenging to formulate and estimate. 
  It also connects to the question of which queue dynamics produce a sticky point in the scaling limit, discussed in Section~\ref{sec:disc}.
\end{enumerate}

\subsubsection{Inference and diagnostics}

\paragraph{Dependence.} Repeated visits to the same level within a trading day may be dependent. 
We use the Andersen--Gill counting-process formulation
\citep{AndersenGill1982}, with a Lin--Wei sandwich variance estimator \citep{LinWei1989} clustered by trading day. Cluster-robust inference relies on a sufficiently large number of clusters. The application includes 250 clusters per stock, whereas the first simulation in \OAsim{} includes twelve and therefore also reports model-based standard errors.

\paragraph{Proportionality.} Level $\times$ log-depth interactions, tested jointly,
check the assumption that Proposition~\ref{prop:nest}(ii) identifies in
\eqref{eq:BO}; rejection calls for reporting depth-specific effects $\gamma_r(q)$ rather than a common effect $\gamma_r$. A
$\gamma_r\times\log u$ interaction, or Schoenfeld residuals, checks proportionality in
duration: a rejection means the stickiness is concentrated in long visits, i.e.\
replenishment rather than initial depth, \emph{provided} the depth coefficient varies
with duration as in \eqref{eq:optd}, since otherwise an entry-depth covariate with a constant coefficient and a
duration-varying effect are observationally similar (\OAladder).

\paragraph{Estimation and diagnostic specifications.} Following Remark~\ref{rem:estimand}, we report $S_r$ from the flat-baseline fit, which targets it directly, and conduct the diagnostics using \eqref{eq:optd}, which provides a more flexible description of duration dependence. The coefficients of the diagnostic model describe conditional hazard contrasts and should be distinguished from the reported estimate of $S_r$.

\paragraph{Functional form and placebos.} 
We compare a natural spline in log depth with a log-linear specification and examine placebo levels, such as prices \$0.03 above each round number, at which no systematic effect is expected. Functional-form specification is particularly important when the level of interest has an unusual depth distribution, since misspecification can distort the estimated level effect (Section~\ref{sec:sim-qr}).

\paragraph{What is not identified.} Depth is chosen by traders partly in anticipation
of how the level will behave. $\gamma_r$ is therefore a descriptive contrast --- how
this level's departure hazard compares with others' at the same visible depth --- and
not the causal effect of a price being round. Causal identification would require exogenous variation in the round-number lattice, potentially arising from tick-size changes, decimalisation, or stock splits.

%==============================================================================

\section{Simulation study}\label{sec:sim}
\label{sec:sim-qr}

\OAsim{} reports a first study in which the departure hazard is imposed directly. It demonstrates the two limitations of simplified specifications discussed in Section~\ref{sec:depth}, but it
cannot test serial dependence in the composition weights, or the endogenous-covariate
problem of Section~\ref{sec:endog}, which is invisible when depth is constant within a
visit. This section extends the experiment by generating price departures from queue dynamics.

\subsection{Design}

Each level carries two best queues evolving as birth--death processes: limit
orders arrive at rate $\lambda$ per side, each resting unit is cancelled at rate
$\theta$, and market orders arrive at rate $\mu$. The price moves \emph{down} when
the bid queue empties and \emph{up} when the ask queue empties, at which point
both queues are redrawn. The departure time is therefore determined by queue depletion, and the simulated mechanisms are specified through the queue dynamics:

\begin{center}
\begin{tabular}{@{}llp{8.0cm}@{}}
\toprule
\textbf{A} & $100.00$ & post-move depth distribution with greater mean depth, \emph{identical}
              per-unit dynamics. Residual stickiness is $1$ by construction \\
\textbf{B} & $99.95$  & limit-order arrival rate $\times 1.65$: the queue
              \emph{replenishes}. Sticky at matched entry depth \\
\textbf{C} & $100.05$ & market-buy rate $\times 1.35$, increased ask-side consumption, producing directional asymmetry and shorter durations \\
\textbf{D} & $\geq 100.10$ & every rate $\times 2$: an exact time change, so the
              hazard doubles at every queue state \\
\bottomrule
\end{tabular}
\end{center}

A liquidity regime $L$, an AR(1) in logs updated at each price move, scales the
post-move depth distribution. Entry depth is consequently strongly persistent ---
its lag-one autocorrelation at a control level is $0.739$ --- so the composition
weights $w_k(\cdot)$ of \eqref{eq:BO} are serially dependent, addressing a limitation of the first design. We simulate $1200$ independent books, yielding $247{,}533$ visits and $5{,}293{,}425$ book updates; the interval-level fits use
the piecewise-exponential form \eqref{eq:poisson} with one row per update.

Because departure times are generated by queue dynamics, the true duration ratios are not available in closed form. We obtain the matched-entry-depth duration ratios from an auxiliary Monte Carlo simulation of $200{,}000$ isolated visits per level, all initialised from a common entry-depth distribution, yielding $S^{\rm true}_B = 1.882\ (\pm 0.005)$,
$S^{\rm true}_C = 0.804$, $S^{\rm true}_D = 0.496\ (\pm 0.001)$, and
$S^{\rm true}_A = 1$ exactly.
 
\subsection{Estimator performance under alternative depth specifications}

\begin{table}[t]
\centering
\small
\begin{tabular}{@{}lrrrrr@{}}
\toprule
& & no depth control & depth bins & \multicolumn{2}{c}{entry depth} \\
\cmidrule(lr){3-3}\cmidrule(lr){4-4}\cmidrule(lr){5-6}
Level & truth & raw $\hat\theta_k/\bar\theta$ & binned \eqref{eq:BO} & log-linear & spline \\
\midrule
\textbf{A} $100.00$ & $1.000$ & $2.332$ \tiny$(0.008)$ & $1.087$ & $0.897$ \tiny$(0.006)$ & $\mathbf{1.000}$ \tiny$(0.005)$ \\
\textbf{B} $99.95$  & $1.882$ & $1.809$ \tiny$(0.015)$ & $1.820$ & $1.872$ \tiny$(0.013)$ & $\mathbf{1.869}$ \tiny$(0.014)$ \\
\textbf{C} $100.05$ & $0.804$ & $0.809$ \tiny$(0.012)$ & $0.806$ & $0.806$ \tiny$(0.008)$ & $\mathbf{0.806}$ \tiny$(0.008)$ \\
\textbf{D} $\geq 100.10$ & $0.496$ & $0.470$ \tiny$(0.032)$ & $0.490$ & $0.495$ \tiny$(0.013)$ & $\mathbf{0.495}$ \tiny$(0.013)$ \\
\bottomrule
\end{tabular}
\caption{Queue-reactive book: residual stickiness index $S_r = e^{-\gamma_r}$
against the matched-entry-depth ground truth. Standard errors clustered by book
($1200$ clusters). The binned column is the residual factor of \eqref{eq:BO} with
eight depth bins.}
\label{tab:qr}
\end{table}

Table~\ref{tab:qr} reports the estimation results.  With entry depth entered as a natural spline, the estimates closely match the true values at all four levels: $1.000$ against $1.000$ at A, $1.869$ against
$1.882$ at B, $0.806$ against $0.804$ at C, $0.495$ against $0.496$ at D. Serial dependence in the composition weights does not invalidate the decomposition in this design because the regime affects departure times entirely through entry depth. This condition would fail if the regime also affected replenishment. The stock, day, and time-of-day strata are intended to account for such variation in the empirical analysis.
The simpler depth specifications produce appreciable bias at level A.

\begin{itemize}
\item The \textbf{binned decomposition} again overestimates residual stickiness at A ($1.087$ versus the true value of $1$) and underestimates it at B. 
\item The \textbf{log-linear depth specification} yields an estimate of $0.897\ (\pm0.006)$ at A, approximately $10\%$ below the true value and seventeen standard errors away, while estimates at B, C, and D remain close to their true values.
\end{itemize}

These results illustrate the sensitivity of the decomposition to the
depth specification. When a level's depth distribution differs markedly
from that of the control levels, coarse or overly restrictive depth
adjustments can introduce systematic bias, as observed at A. The magnitude
of the log-linear bias relative to its standard error indicates that
sampling uncertainty does not explain the discrepancy. The raw estimates
also show that the composition effect operates in both directions:
it is large and positive at A (a raw ratio of $2.33$ compared with a
residual of $1.00$), but negative at D. In the latter case, low-liquidity
regimes permit larger price excursions, so levels ten or more ticks above
$100.00$ are visited disproportionately when queues are thin.

The direction equation also recovers the imposed asymmetry. At level C, the estimated log-odds shift is $+0.608\ (\pm0.022)$, corresponding to an upward-departure probability of $0.5989$, compared with the matched-depth true value of $0.5942$.

\subsection{The depth covariate's lag}

\begin{figure}[t]
\centering
\includegraphics[width=\textwidth]{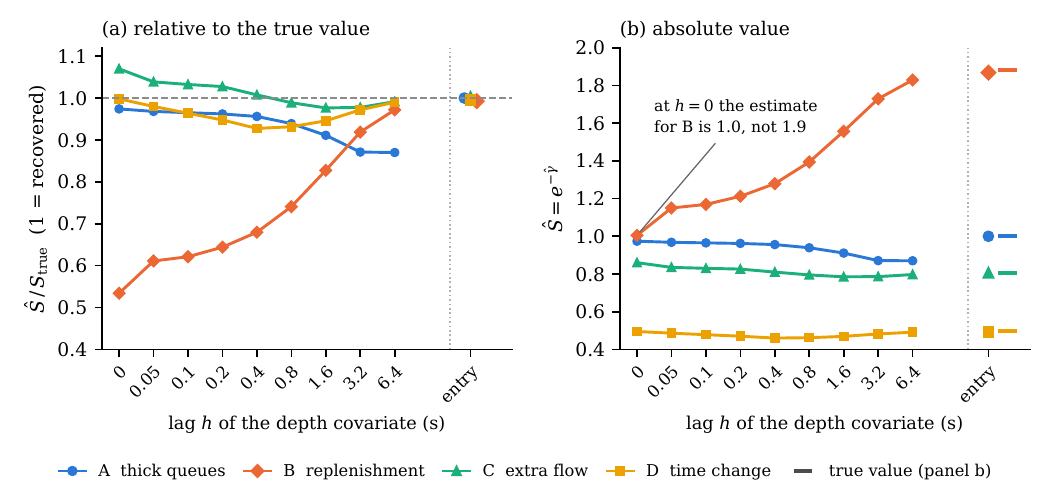}
\caption{\caplag}
\label{fig:lag}
\end{figure}

Figure~\ref{fig:lag} illustrates the covariate-timing issue discussed in Section~\ref{sec:endog}. 
We refit the same interval-level model with the depth covariate measured $h$ seconds
before the current instant, for each of the four simulated mechanisms: 

\begin{itemize}
\item At $h=0$, level B's estimated residual stickiness is $1.006$,
  compared with the matched-entry-depth true value of $1.882$.
  Conditioning on the contemporaneous queue state largely absorbs
  the persistence associated with replenishment.
\item The estimates at C and D are comparatively stable over the
  lags examined ($0.86 \to 0.81$ and $0.50 \to 0.50$, respectively).
  Unlike replenishment at B, these mechanisms alter transition rates
  conditional on the queue state.
\item Intermediate lags change the conditioning information and
  should not be interpreted as robustness checks for a common
  estimand. The resulting profile need not be monotone, as illustrated
  by level A.
\end{itemize}

Within this simulation design, the lag profile distinguishes the
replenishment mechanism at B from the increased order-flow and
time-scaling mechanisms at C and D. It therefore provides a mechanism
diagnostic rather than a robustness check for a fixed estimand.

% \subsection{Duration dependence, and the limits of entry depth}
Duration dependence provides further evidence on the limitations of entry-depth adjustment. The test of proportional hazards rejects the null ($\chi^2_4=5268$). The coefficient on the interaction between level and elapsed time is $-0.367\ (\pm0.007)$ at B, consistent with the replenishment mechanism discussed in Section~\ref{sec:ph}, and $+0.007\ (\pm0.009)$ at D, consistent with a pure time change. At A, however, the coefficient is $+0.381\ (\pm0.009)$ despite the absence of a residual effect: entry depth becomes less informative about the current queue state as a visit lengthens, and the level indicator captures part of this discrepancy. Allowing the depth coefficient to vary with elapsed time substantially reduces this artefact while preserving the effect at B (\OAladder).

%==============================================================================
\section{Application}\label{sec:app}
%==============================================================================
 
% This section applies the framework to the five stocks described in Section~\ref{sec:data}. We examine whether prices remain longer at round levels after conditioning on visible liquidity, and whether departure directions differ between round and nonround levels. 

\begin{table}[t]
\centering
\footnotesize
\setlength{\tabcolsep}{4pt}
\resizebox{\textwidth}{!}{%
\begin{tabular}{@{}lrrrrr@{}}
\toprule
& Apple & Cisco & Intel & SoFi & Am.\ Airlines \\
\midrule
trading days & 250 & 250 & 250 & 250 & 250 \\
price range, 1st--99th percentile of visits (\$) & 191--281 & 53--80 & 19--43 & 10--32 & 9--19 \\
median price (\$) & 215.10 & 67.35 & 32.12 & 26.26 & 12.69 \\
tick as a fraction of the median price (bp) & 0.5 & 1.5 & 3.1 & 3.8 & 7.9 \\
session time with a one-tick spread (\%) & 28.7 & 93.7 & 97.5 & 96.1 & 99.3 \\
session time retained after the wide-spread exclusion (\%) & 66.8 & 99.2 & 99.97 & 99.86 & 99.99 \\
visits with an observed entry and positive duration & 4,363,496 & 1,603,601 & 1,635,439 & 1,376,113 & 291,109 \\
\quad of which censored (\%) & 11.8 & 1.1 & 0.3 & 0.6 & 0.3 \\
\quad of which ending in a multi-tick move (\%) & 7.5 & 1.4 & 0.8 & 0.8 & 0.3 \\
visits with a round dollar as best bid (thousands) & 50.1 & 15.4 & 15.8 & 13.2 & 2.7 \\
visits with a round dollar as best ask (thousands) & 50.9 & 15.3 & 15.8 & 13.4 & 2.7 \\
mean visit duration, round dollar / other levels (s) & 0.76 / 0.79 & 3.63 / 3.60 & 3.24 / 3.56 & 4.35 / 4.24 & 18.3 / 20.1 \\
median visit duration, round dollar / other levels (s) & 0.16 / 0.21 & 0.67 / 0.73 & 0.48 / 0.60 & 0.82 / 0.92 & 6.0 / 6.7 \\
\bottomrule
\end{tabular}} 
\caption{Sample characteristics, ordered by the size of a one-cent tick relative to the median visit price (in basis points). Visits are defined in Section~\ref{sec:visits}. The first visit of each day and the first visit after an observation interruption are excluded because their entry times are not observed. ``Other levels'' are levels assigned to none of the round-number categories.}
\label{tab:sample}
\end{table}

Table~\ref{tab:sample} summarises the five samples, ordered by relative tick size. Apple differs markedly from the other stocks: its spread equals one tick during only 28.7\% of session time, compared with 93.7--99.3\% for the other four stocks. The spread filters retain 66.8\% of Apple's session time, compared with more than 99\% for the others. Apple also has higher proportions of censored visits and visits ending in multi-tick moves. American Airlines has substantially fewer visits: approximately 291,000, compared with 1.38--4.36 million for the other stocks. Its round-dollar visit counts are correspondingly smaller, limiting the precision of estimates at those levels.

\subsection{Specification}\label{sec:app-spec}

The estimand is the residual stickiness index $S_r$ defined in \eqref{eq:estimands}: the ratio of mean visit durations at matched entry depth for level type $r$ relative to control levels. Following Remark~\ref{rem:estimand}, we estimate this index using the flat-baseline specification of \eqref{eq:poisson}, with $\hat S_r$ given by the reciprocal of the estimated departure hazard ratio for level type $r$. A value below one indicates shorter mean visit durations relative to the control levels. The specification includes the following:

\begin{itemize}
\item \textbf{Depth} $Z_i(0)$: log displayed depth at entry on the bid and ask sides, each represented by a natural cubic spline with six degrees of freedom and knots placed at quantiles, allowing for the nonlinear depth effects illustrated in Section~\ref{sec:sim-qr};
\item \textbf{Strata} $s(i)$: day-by-half-hour-of-entry strata, with 3{,}232 strata per stock, allowing baseline departure rates to vary across days and intraday intervals;
\item \textbf{Level types} $R_{k(i)}$: mutually exclusive round-dollar, half-dollar, quarter-dollar, and dime categories, each represented separately according to whether the round price is the best bid or best ask, giving eight indicators;
\item \textbf{Controls}: distance in ticks to the nearest round dollar, grouped into bands (up to 2, 3--5, 6--10, 11--20, 21--35, and over 35), to account for proximity to round-dollar prices; signed distance from the day's opening price, represented by a natural spline with four degrees of freedom, to account for drift (Proposition~\ref{prop:drift}); indicators for a quote within one tick of the previous closing price and the day's opening price; and the direction and size of the price move initiating the visit;
\item \textbf{Inference}: day-clustered sandwich standard errors (250 clusters), $t_{249}$ critical values, and Benjamini--Hochberg adjustment across the eight round-number types within each stock.
\end{itemize}

We estimate the direction equation \eqref{eq:ph-dir} using logistic regression on single-tick departures, with an upward departure as the outcome, day effects, and the same covariates and level-type indicators. $\hat\alpha_r$ denotes the unadjusted share of upward departures, and $\hat\phi_r$ the adjusted log-odds difference relative to the control levels. Placebo indicators for prices three cents above a round dollar (Section~\ref{sec:levels}) are included in both equations. 
The same specification is applied to all five stocks.

\begin{table}[t]
\centering
\footnotesize
\begin{tabular}{@{}lrrrcrl@{}}
\toprule
& & \multicolumn{3}{c}{stickiness} & \multicolumn{2}{c}{direction} \\
\cmidrule(lr){3-5}\cmidrule(lr){6-7}
Level type & visits & raw & composition & residual $\hat S_r$ [95\% CI] & $\hat\alpha_r$ & \phantom{$+$ }$\hat\phi_r$ \phantom{$+$}[95\% CI] \\
\midrule
\multicolumn{7}{@{}l}{\emph{Apple}} \\
round dollar, best bid & 50,142 & 0.94 & 0.99 & 0.94\; [0.92, 0.96] & 0.50 & $-$0.02\; [$-$0.05, 0.01] \\
round dollar, best ask & 50,919 & 0.97 & 0.98 & 0.99\; [0.97, 1.02] & 0.48 & \phantom{$+$}0.01\; [$-$0.02, 0.04] \\
\addlinespace[2pt]
placebo, best bid & 45,694 & 0.99 & 1.00 & 0.99\; [0.97, 1.01] & 0.49 & $-$0.01\; [$-$0.04, 0.01] \\
placebo, best ask & 46,555 & 1.01 & 1.00 & 1.01\; [0.99, 1.03] & 0.49 & $-$0.02\; [$-$0.05, 0.01] \\
\midrule
\multicolumn{7}{@{}l}{\emph{Cisco}} \\
round dollar, best bid & 15,410 & 0.98 & 1.00 & 0.98\; [0.94, 1.01] & 0.51 & \phantom{$+$}0.02\; [$-$0.03, 0.07] \\
round dollar, best ask & 15,283 & 0.99 & 1.00 & 0.99\; [0.95, 1.03] & 0.49 & $-$0.02\; [$-$0.07, 0.03] \\
\addlinespace[2pt]
placebo, best bid & 16,401 & 0.99 & 1.00 & 0.99\; [0.95, 1.03] & 0.50 & \phantom{$+$}0.00\; [$-$0.05, 0.05] \\
placebo, best ask & 16,327 & 1.00 & 1.00 & 1.00\; [0.97, 1.04] & 0.50 & $-$0.01\; [$-$0.06, 0.04] \\
\midrule
\multicolumn{7}{@{}l}{\emph{Intel}} \\
round dollar, best bid & 15,829 & 0.96 & 1.02 & 0.94\; [0.90, 0.98] & 0.54 & \phantom{$+$}0.11\; [0.05, 0.16] \\
round dollar, best ask & 15,790 & 0.94 & 1.01 & 0.93\; [0.89, 0.97] & 0.46 & $-$0.09\; [$-$0.15, $-$0.03] \\
\addlinespace[2pt]
placebo, best bid & 16,538 & 1.02 & 0.99 & 1.03\; [0.99, 1.06] & 0.50 & \phantom{$+$}0.00\; [$-$0.05, 0.04] \\
placebo, best ask & 16,535 & 0.99 & 0.99 & 1.01\; [0.97, 1.05] & 0.50 & \phantom{$+$}0.00\; [$-$0.04, 0.04] \\
\midrule
\multicolumn{7}{@{}l}{\emph{SoFi}} \\
round dollar, best bid & 13,169 & 0.94 & 1.01 & 0.93\; [0.90, 0.96] & 0.53 & \phantom{$+$}0.06\; [0.00, 0.11] \\
round dollar, best ask & 13,398 & 0.94 & 1.01 & 0.93\; [0.90, 0.96] & 0.46 & $-$0.08\; [$-$0.13, $-$0.04] \\
\addlinespace[2pt]
placebo, best bid & 14,250 & 1.01 & 0.99 & 1.01\; [0.98, 1.05] & 0.49 & $-$0.02\; [$-$0.06, 0.02] \\
placebo, best ask & 14,304 & 0.97 & 0.99 & 0.98\; [0.95, 1.01] & 0.51 & \phantom{$+$}0.02\; [$-$0.03, 0.07] \\
\midrule
\multicolumn{7}{@{}l}{\emph{American Airlines}} \\
round dollar, best bid & 2,712 & 0.95 & 1.01 & 0.94\; [0.87, 1.02] & 0.51 & \phantom{$+$}0.04\; [$-$0.07, 0.15] \\
round dollar, best ask & 2,737 & 1.04 & 1.01 & 1.03\; [0.95, 1.12] & 0.48 & $-$0.02\; [$-$0.15, 0.10] \\
\addlinespace[2pt]
placebo, best bid & 2,531 & 0.99 & 0.99 & 1.00\; [0.92, 1.09] & 0.51 & \phantom{$+$}0.03\; [$-$0.06, 0.12] \\
placebo, best ask & 2,527 & 0.92 & 1.00 & 0.93\; [0.87, 0.99] & 0.49 & $-$0.03\; [$-$0.14, 0.08] \\
\bottomrule
\end{tabular} 
\caption{Residual stickiness and departure direction at round-dollar levels, 2025. The raw duration index is estimated without depth splines, while the residual index $\hat S_r$ includes them; all other controls are identical across the two specifications. The composition factor is the raw index divided by the residual index, following the multiplicative decomposition in \eqref{eq:BO}. An index below one indicates shorter mean visit durations relative to the control levels. $\hat\alpha_r$ is the unadjusted share of single-tick departures that are upward, and $\hat\phi_r$ is the adjusted log-odds difference relative to the control levels. Bid and ask rows distinguish whether the round price is the best bid or best ask of the visit (Section~\ref{sec:levels}). Placebo levels are three cents above round-dollar prices. Confidence intervals use day-clustered standard errors. Results for half-dollar, quarter-dollar, and dime levels from the same specifications are reported in \OAtabfull{}. Benjamini--Hochberg adjustment is applied across the eight round-number types within each stock. Adjusted $p$-values for the round-dollar residual indices are: Apple $<0.001$ (bid) and $0.50$ (ask); Cisco $0.50$ and $0.58$; Intel $0.01$ and $0.004$; SoFi $<0.001$ and $<0.001$; American Airlines $0.33$ and $0.73$.}

\label{tab:app-main}
\end{table}

\subsection{Visit durations and departure directions at round levels}

\paragraph{Visit durations.}
Table~\ref{tab:app-main} provides no evidence of longer depth-adjusted
durations at round-dollar levels in any of the five stocks. Of the ten
round-dollar indices, nine are below one; the remaining estimate,
$1.03$ for American Airlines on the ask side, is not statistically
distinguishable from one. No half-dollar, quarter-dollar, or dime index
is significantly above one either (\OAtabfull{}). The descriptive
statistics show a similar pattern before conditioning: median visit
durations at round-dollar levels are shorter than at other levels in
all five stocks, while mean durations are shorter in three stocks and
differ by less than three per cent in Cisco and SoFi
(Table~\ref{tab:sample}).

All statistically significant round-dollar duration indices are below
one. In Intel, the estimates are $0.94$ $[0.90, 0.98]$ at a round-dollar
bid and $0.93$ $[0.89, 0.97]$ at a round-dollar ask. In SoFi, both
estimates are $0.93$ $[0.90, 0.96]$. These estimates indicate mean
durations approximately six to seven per cent shorter than at
comparable control levels. The effects remain significant after
multiplicity adjustment, while the corresponding placebo indices
range from $0.98$ to $1.03$. Apple's estimates are $0.94$ $[0.92, 0.96]$
on the bid side and $0.99$ $[0.97, 1.02]$ on the ask side. Neither
Cisco nor American Airlines shows a statistically significant
round-dollar duration difference.

Apple's duration reduction is confined to the bid side and becomes
progressively smaller across half-dollar, quarter-dollar, and dime
levels (\OAtabfull{}). Restricting exposure to time with a one-tick
spread moves the round-dollar bid index to $0.96$ and the ask index
above one. This sensitivity indicates that the treatment of time
spent at wider spreads contributes to the estimated pattern. Given
that the observation filters also exclude approximately one-third
of Apple's session time, we do not pursue the interpretation of
this duration effect further.

\paragraph{Departure directions.}
The direction estimates reveal an additional pattern in Intel and
SoFi: departures are biased away from round-dollar levels. In Intel,
54 per cent of single-tick departures are upward when the round dollar
is the best bid, compared with 46 per cent when it is the best ask.
The corresponding adjusted log-odds differences are $0.11$
$[0.05, 0.16]$ and $-0.09$ $[-0.15, -0.03]$. SoFi exhibits the same
pattern, with estimates of $0.06$ $[0.00, 0.11]$ and $-0.08$
$[-0.13, -0.04]$. An upward departure when the round dollar is the
best bid moves the quote pair away from that price, as does a downward
departure when it is the best ask. Neither stock exhibits a
statistically significant directional effect at the placebo levels.
 
\begin{figure}[t]
\centering
\includegraphics[width=\textwidth]{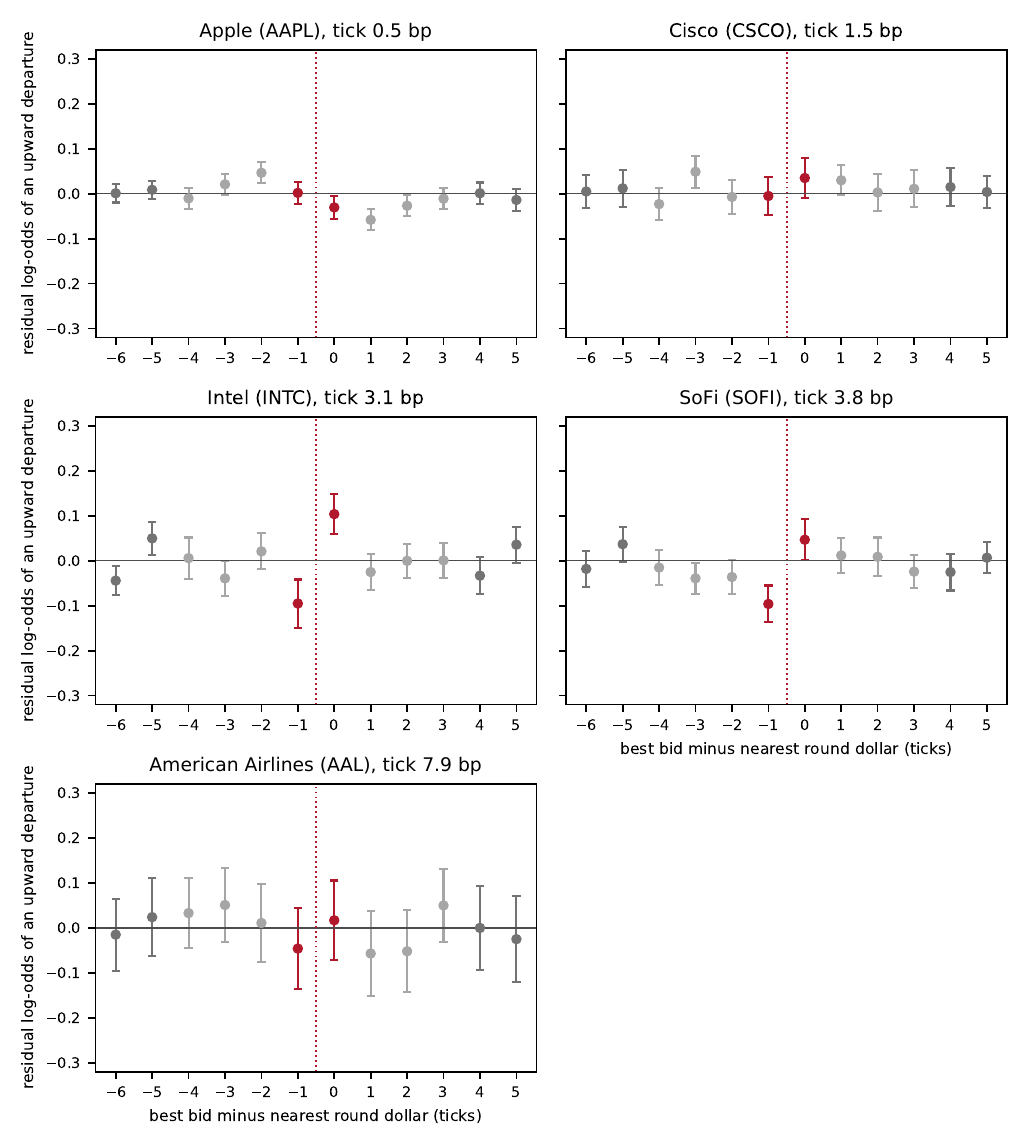}
\caption{\capprofile}
\label{fig:profile}
\end{figure}

Figure~\ref{fig:profile} shows that this directional asymmetry is
localised around round-dollar levels. We replace the round-dollar
indicators with indicators for each position of the best bid from six
ticks below to five ticks above the nearest round dollar. In Intel
and SoFi, the largest adjusted log-odds differences occur at the two
positions corresponding to a round-dollar bid or ask, where they
have opposite signs. Their magnitudes are approximately $0.10$,
except for SoFi's bid-side estimate of approximately $0.05$. In Intel,
the nickel positions five cents away exhibit a similar pattern at
approximately half the magnitude. The localised deviations resemble
the isolated skew level of the first simulation (\OAsim), rather than the smooth spatial
variation associated with drift (Proposition~\ref{prop:drift}).

The other three stocks show no statistically significant
round-dollar directional effect. In Cisco and Apple, the adjusted
log-odds differences are within $0.03$ of zero. Apple's confidence
intervals, $[-0.05, 0.01]$ on the bid side and $[-0.02, 0.04]$ on the
ask side, exclude the corresponding point estimates for Intel and
SoFi. The spatial profiles fluctuate around zero, apart from a small
increase two ticks below the round dollar and a small decrease one
tick above it in Apple. These deviations appear in both halves of
the sample but do not occur at the round-dollar bid or ask positions.

American Airlines has adjusted log-odds differences of $0.04$
$[-0.07, 0.15]$ on the bid side and $-0.02$ $[-0.15, 0.10]$ on the
ask side. These intervals contain both zero and the corresponding
Intel and SoFi estimates. Its substantially smaller visit count
limits precision: the estimated probability of detecting an effect
of Intel's magnitude at the five per cent level is approximately
one half on the bid side and one third on the ask side. Its
directional results are therefore inconclusive.

\begin{figure}[t]
\centering
\includegraphics[width=\textwidth]{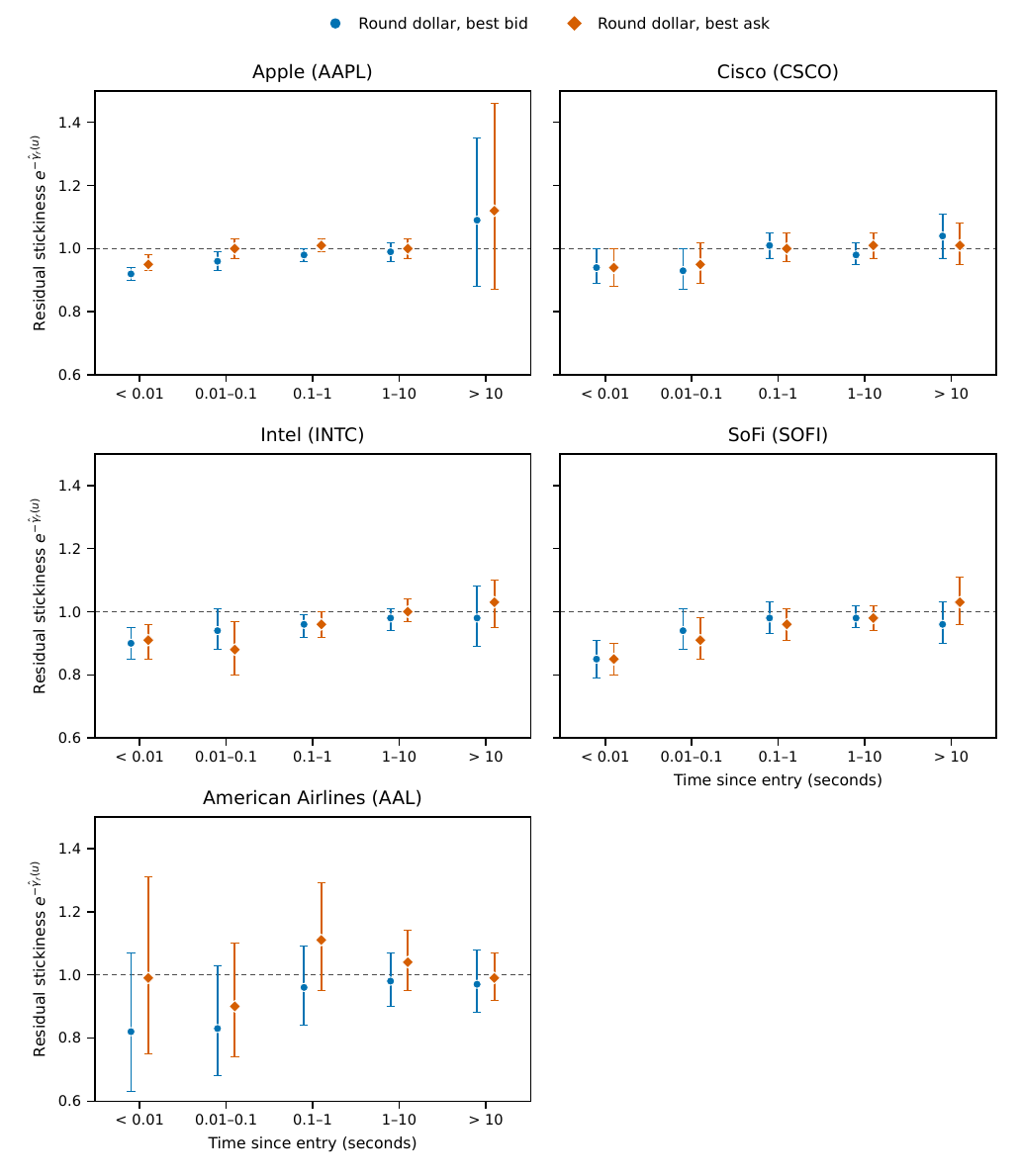}
\caption{\capelapsed}
\label{fig:elapsed}
\end{figure}

\paragraph{Temporal variation.} 
Table~\ref{tab:robust} and Figure~\ref{fig:elapsed} show that
departure-rate differences are concentrated shortly after entry.
During the first $0.01$ seconds, inverse hazard ratios are
$0.90$--$0.91$ for Intel and $0.85$ for SoFi.
The estimates subsequently approach one: neither side differs
significantly from one beyond one second for Intel or beyond
$0.1$ seconds for SoFi. Cisco shows a weaker early pattern,
with confidence intervals containing one throughout
(\OAtabrobust{}).

Across intraday periods, the reductions in mean duration are most
evident during the first trading hour. The bid- and ask-side duration
indices are $0.94$ and $0.87$ in Intel and $0.89$ on both sides in
SoFi (Table~\ref{tab:robust}). A test of equality across quarterly coefficients in Intel does not
reject the null ($p=0.88$ on the bid side and $p=0.10$ on the ask side).

\subsection{Liquidity concentration and composition effects}

Displayed liquidity concentrates at round-dollar levels in all five
stocks. Within day-by-half-hour strata, displayed depth at round-dollar
quotes exceeds that at other levels. The bid- and ask-side depth ratios
are $1.07$ $[1.04, 1.11]$ and $1.12$ $[1.07, 1.16]$ in Cisco,
$1.28$ and $1.37$ in Apple, $1.24$ and $1.28$ in American Airlines,
$1.47$ on both sides in Intel, and $1.50$ and $1.46$ in SoFi.
All confidence intervals exclude one. Liquidity concentration is
therefore present throughout the sample and is strongest in Intel
and SoFi, the two stocks with directional asymmetry at round-dollar
levels.

Despite these depth differences, the composition factors in
Table~\ref{tab:app-main} range from $0.98$ to $1.02$ across all
round-dollar categories. In Intel and SoFi, they are between $1.01$
and $1.02$, indicating a small positive contribution of depth
composition to relative duration. Adding depth controls changes
Intel's round-dollar duration index from $0.96$ to $0.94$ on the
bid side and from $0.94$ to $0.93$ on the ask side; in SoFi, both
indices move from $0.94$ to $0.93$. Thus, shorter durations are
already apparent before depth adjustment, and the estimated
reductions become slightly larger after conditioning on depth.
The greater displayed liquidity at round-dollar levels does not
account for these shorter durations.

\subsection{Robustness}

Panel A of Table~\ref{tab:robust} reports the round-dollar duration
indices for Intel and SoFi under alternative specifications and
sample restrictions. Corresponding results for the other three
stocks are reported in \OAtabrobust{}. We vary the depth and distance
controls, restrict visits to within five cents of a round dollar or
to common depth support, and retain only single-tick entries.
Additional checks exclude the first and last five minutes of trading
or the week of 7 April 2025, replace day-by-half-hour strata with
additive effects, measure exposure using only time with a one-tick
spread, and treat spread-limit exits as departures rather than
censoring events.

In SoFi, the duration indices remain between $0.93$ and $0.96$
across all specifications and sample restrictions in Panel A,
and all confidence intervals exclude one. Intel's point estimates
also remain below one, although statistical significance depends
on the specification. The bid-side confidence interval includes
one when depth controls are omitted, when distance bands are
replaced by a smooth function, and when the sample is restricted
to common depth support. The ask-side interval also includes one
under the smooth distance specification. Omitting distance controls
produces intervals containing one on both sides, indicating
sensitivity to how proximity to round-dollar levels is accounted for.

Among the alternative specifications for Cisco, only one
smooth-distance estimate differs significantly from one. In
American Airlines, four specifications yield bid-side indices
between $0.92$ and $0.93$ with confidence intervals excluding one;
none of the ask-side intervals excludes one. Apple's bid-side
indices range from $0.92$ to $0.96$, while its ask-side indices
range from $0.97$ to $1.03$. Across the ten stock-by-side placebo
estimates in Table~\ref{tab:app-main}, only the American Airlines
ask-side estimate, $0.93$ $[0.87, 0.99]$, has a confidence interval
excluding one.

We also reconstruct Cisco's price series using stricter and more
permissive spread filters than those in Section~\ref{sec:cleaning}.
Tightening the cutoffs to two ticks and thirty seconds, or relaxing
them to five ticks and 120 seconds, changes the round-dollar duration
indices by no more than $0.005$. Under the stricter filters, the
proportion of censored visits increases from approximately one
per cent to five per cent.

\begin{table}[t]
\centering
\footnotesize
\setlength{\tabcolsep}{4pt}
\renewcommand{\arraystretch}{1.05}

% Default: invisible dagger to preserve alignment.
% Use \robci[\dagger]{estimate}{lower}{upper} for marked cells.
\newcommand{\robci}[4][\phantom{\dagger}]{%
  \ensuremath{#2\,[#3,#4]^{#1}}}

\begin{tabular}{@{}lcccc@{}}
\toprule
& \multicolumn{2}{c}{Intel}
& \multicolumn{2}{c}{SoFi} \\
\cmidrule(lr){2-3}\cmidrule(lr){4-5}
Specification / subsample
& Best bid & Best ask & Best bid & Best ask \\

\midrule
\multicolumn{5}{@{}l}{\textbf{Panel A. Robustness checks}} \\
\addlinespace[3pt]

Primary specification
& \robci[\dagger]{0.94}{0.90}{0.98}
& \robci[\dagger]{0.93}{0.89}{0.97}
& \robci[\dagger]{0.93}{0.90}{0.96}
& \robci[\dagger]{0.93}{0.90}{0.96} \\

\addlinespace[3pt]
\multicolumn{5}{@{}l}{Model specification} \\
\addlinespace[1pt]

\quad No depth control
& \robci{0.96}{0.92}{1.00}
& \robci[\dagger]{0.94}{0.91}{0.98}
& \robci[\dagger]{0.94}{0.91}{0.97}
& \robci[\dagger]{0.94}{0.91}{0.97} \\

\quad Log-linear depth
& \robci[\dagger]{0.94}{0.90}{0.98}
& \robci[\dagger]{0.93}{0.89}{0.97}
& \robci[\dagger]{0.93}{0.90}{0.96}
& \robci[\dagger]{0.93}{0.90}{0.96} \\

\quad No distance bands
& \robci{0.95}{0.89}{1.02}
& \robci{0.95}{0.89}{1.00}
& \robci[\dagger]{0.94}{0.91}{0.98}
& \robci[\dagger]{0.94}{0.91}{0.98} \\

\quad Smooth distance control
& \robci{0.97}{0.92}{1.02}
& \robci{0.96}{0.91}{1.01}
& \robci[\dagger]{0.95}{0.92}{0.98}
& \robci[\dagger]{0.95}{0.92}{0.98} \\

\quad Placebo indicators added
& \robci[\dagger]{0.94}{0.91}{0.98}
& \robci[\dagger]{0.93}{0.90}{0.97}
& \robci[\dagger]{0.93}{0.90}{0.96}
& \robci[\dagger]{0.93}{0.90}{0.95} \\

\quad Additive day/half-hour effects
& \robci[\dagger]{0.93}{0.90}{0.97}
& \robci[\dagger]{0.94}{0.90}{0.98}
& \robci[\dagger]{0.93}{0.90}{0.97}
& \robci[\dagger]{0.95}{0.91}{0.98} \\

\addlinespace[3pt]
\multicolumn{5}{@{}l}{Sample and event definition} \\
\addlinespace[1pt]

\quad Within $\pm5$ cents (matched)
& \robci[\dagger]{0.94}{0.91}{0.98}
& \robci[\dagger]{0.96}{0.92}{1.00}
& \robci[\dagger]{0.94}{0.91}{0.97}
& \robci[\dagger]{0.93}{0.90}{0.95} \\

\quad Single-tick entries only
& \robci[\dagger]{0.94}{0.90}{0.99}
& \robci[\dagger]{0.93}{0.90}{0.97}
& \robci[\dagger]{0.93}{0.91}{0.96}
& \robci[\dagger]{0.93}{0.90}{0.96} \\

\quad Common depth support
& \robci{0.96}{0.92}{1.01}
& \robci[\dagger]{0.95}{0.91}{0.99}
& \robci[\dagger]{0.96}{0.93}{0.99}
& \robci[\dagger]{0.95}{0.91}{0.98} \\

\quad Exclude first/last 5 min
& \robci[\dagger]{0.95}{0.91}{0.99}
& \robci[\dagger]{0.94}{0.90}{0.98}
& \robci[\dagger]{0.94}{0.91}{0.98}
& \robci[\dagger]{0.94}{0.91}{0.97} \\

\quad Exclude week of 7 April
& \robci[\dagger]{0.95}{0.90}{0.99}
& \robci[\dagger]{0.94}{0.90}{0.98}
& \robci[\dagger]{0.93}{0.90}{0.96}
& \robci[\dagger]{0.93}{0.90}{0.96} \\

\quad Exposure in one-tick time
& \robci[\dagger]{0.94}{0.90}{0.98}
& \robci[\dagger]{0.93}{0.89}{0.98}
& \robci[\dagger]{0.94}{0.91}{0.97}
& \robci[\dagger]{0.94}{0.90}{0.97} \\

\quad Spread-limit exit as event
& \robci[\dagger]{0.94}{0.90}{0.98}
& \robci[\dagger]{0.93}{0.89}{0.97}
& \robci[\dagger]{0.93}{0.90}{0.96}
& \robci[\dagger]{0.93}{0.90}{0.96} \\

\midrule
\multicolumn{5}{@{}l}{\textbf{Panel B. Temporal variation}} \\
\addlinespace[4pt]

\multicolumn{5}{@{}l}{Time of day} \\
\addlinespace[1pt]

\quad First hour
& \robci[\dagger]{0.94}{0.90}{0.98}
& \robci[\dagger]{0.87}{0.81}{0.93}
& \robci[\dagger]{0.89}{0.84}{0.93}
& \robci[\dagger]{0.89}{0.84}{0.93} \\

\quad 10:30--15:00
& \robci{0.95}{0.89}{1.01}
& \robci{0.99}{0.94}{1.06}
& \robci{0.96}{0.92}{1.01}
& \robci{0.98}{0.93}{1.02} \\

\quad Last hour
& \robci{0.95}{0.82}{1.08}
& \robci{1.01}{0.89}{1.14}
& \robci{0.97}{0.85}{1.10}
& \robci{0.89}{0.79}{1.01} \\

\addlinespace[3pt]
\multicolumn{5}{@{}l}{Calendar quarter, 2025} \\
\addlinespace[1pt]

\quad Q1
& \robci{0.95}{0.89}{1.00}
& \robci{0.97}{0.89}{1.06}
& \robci{0.98}{0.90}{1.07}
& \robci{0.96}{0.90}{1.03} \\

\quad Q2
& \robci{0.99}{0.84}{1.17}
& \robci[\dagger]{0.87}{0.78}{0.96}
& \robci{0.93}{0.84}{1.03}
& \robci{0.95}{0.84}{1.06} \\

\quad Q3
& \robci{0.93}{0.85}{1.02}
& \robci[\dagger]{0.86}{0.76}{0.97}
& \robci[\dagger]{0.92}{0.87}{0.97}
& \robci[\dagger]{0.89}{0.84}{0.94} \\

\quad Q4
& \robci[\dagger]{0.93}{0.88}{0.98}
& \robci{0.97}{0.92}{1.03}
& \robci[\dagger]{0.94}{0.89}{0.98}
& \robci{0.96}{0.91}{1.00} \\

\addlinespace[3pt]
\multicolumn{5}{@{}l}{%
  Time since entry (piecewise exponential)} \\
\addlinespace[1pt]

\quad $<0.01$\,s
& \robci[\dagger]{0.90}{0.85}{0.95}
& \robci[\dagger]{0.91}{0.85}{0.96}
& \robci[\dagger]{0.85}{0.79}{0.91}
& \robci[\dagger]{0.85}{0.80}{0.90} \\

\quad $0.01$--$0.1$\,s
& \robci{0.94}{0.88}{1.01}
& \robci[\dagger]{0.88}{0.80}{0.97}
& \robci{0.94}{0.88}{1.01}
& \robci[\dagger]{0.91}{0.85}{0.98} \\

\quad $0.1$--$1$\,s
& \robci[\dagger]{0.96}{0.92}{0.99}
& \robci{0.96}{0.92}{1.00}
& \robci{0.98}{0.93}{1.03}
& \robci{0.96}{0.91}{1.01} \\

\quad $1$--$10$\,s
& \robci{0.98}{0.94}{1.01}
& \robci{1.00}{0.97}{1.04}
& \robci{0.98}{0.95}{1.02}
& \robci{0.98}{0.94}{1.02} \\

\quad $>10$\,s
& \robci{0.98}{0.89}{1.08}
& \robci{1.03}{0.95}{1.10}
& \robci{0.96}{0.90}{1.03}
& \robci{1.03}{0.96}{1.11} \\

\bottomrule
\end{tabular}

\caption{The round-dollar residual index $\hat S$ for the two
stocks with an effect, under alternative specifications and
samples, with 95\% intervals clustered on day.
A dagger ($\dagger$) marks an interval that excludes one before
rounding.
``No distance bands'' and ``smooth distance'' change the control
for the distance from the nearest round dollar;
``within $\pm5$ cents'' keeps visits within five ticks of a round
dollar and stratifies on day, half-hour, nearest dollar and entry
direction.
The time-of-day, quarter and elapsed-time rows are separate fits.
The same rows for Apple, Cisco and American Airlines are in
\OAtabrobust{}.}
\label{tab:robust}
\end{table}

\subsection{Secondary analyses}

\paragraph{Graded roundness.}
We examine whether duration and directional effects vary with the
degree of roundness. In Intel, both effects are generally stronger
at coarser round-price increments (\OAtabfull{}), with half-dollar
effects comparable to those at whole-dollar levels. The smaller
directional effects at nickel positions in Figure~\ref{fig:profile}
extend this pattern to five-cent increments. In SoFi, the half-dollar
duration index is $0.95$ on the bid side but $0.99$ on the ask side,
while the dime indices are $0.98$ on both sides.
 
\paragraph{Tick size.}
The cross-sectional results suggest a possible association between
relative tick size and directional asymmetry at round-dollar levels.
Apple and Cisco, with relative tick sizes of approximately $0.5$
and $1.5$ basis points, show no significant directional effect,
whereas Intel and SoFi, at approximately three and four basis
points, do. American Airlines has the largest relative tick size,
approximately eight basis points, but its estimates are too
imprecise to establish a directional effect. This pattern does
not establish a causal role for tick size. The stocks also differ
in trading activity, volatility, and market participants, and
two stocks with significant directional effects are insufficient
to distinguish among these explanations.

We therefore examine within-stock variation in relative tick size
for Intel and SoFi. In both the duration and direction equations,
we interact each level-type indicator with the one-cent tick
divided by the entry midpoint, expressed in basis points.
For SoFi, relative tick size ranges from $3.3$ to $8.5$ basis
points between the 5th and 95th percentiles of visits. None of
the four round-dollar interaction coefficients differs
significantly from zero. The direction coefficients are $-0.001$
$[-0.027, 0.025]$ on the bid side and $-0.015$ $[-0.037, 0.007]$
on the ask side per basis point. For Intel, relative tick size
ranges from $2.4$ to $5.1$ basis points. Its duration interaction
coefficients are also indistinguishable from zero, while the
direction coefficients are $-0.059$ $[-0.105, -0.012]$ on the bid
side and $0.050$ $[-0.003, 0.103]$ on the ask side. Both signs
are opposite to those predicted by stronger departures away
from round-dollar levels at larger relative tick sizes.

Both stock prices rose over the year, so the interaction estimates
may also reflect changes in market conditions over time. The
cross-sectional pattern is consistent with a relative-tick-size
explanation, but the within-stock results do not support stronger
directional effects at larger relative tick sizes.

% ==============================================================================
\section{Discussion and conclusion}\label{sec:disc}
%==============================================================================

This paper develops a framework for distinguishing persistence
associated with visible liquidity from additional persistence at
round price levels. Greater displayed depth and longer residence
times alone do not establish whether round levels exhibit such
additional persistence; the relevant comparison is conditional on
liquidity. Modelling visit durations and departure directions
separately also distinguishes prolonged residence from a tendency
for prices to move away from a round level.

Across five Nasdaq stocks in 2025, round-dollar levels attract
greater displayed depth but show no evidence of longer
depth-adjusted residence. In Intel and SoFi, visits are shorter
and departures are biased away from the round number, consistent
with directional barrier effects without additional stickiness.
Differences in departure rates are concentrated shortly after
entry, although this does not establish when the directional
asymmetry arises. Cisco and Apple show no statistically
significant round-dollar directional effect, while estimates
for American Airlines are too imprecise to exclude effects
of the magnitude observed in Intel and SoFi.

The methodological contribution is an occupation measure indexed
jointly by price level and queue state. On a one-tick grid, the
scaled departure count provides a discrete analogue of local time,
and the departure rate multiplied by the squared tick size yields
level-wise volatility. The queue-depletion interpretation connects
these quantities to queue dynamics. Simulations demonstrate the
importance of flexible depth adjustment and of distinguishing
mean-duration ratios from inverse hazard ratios when the duration
baseline is not flat. They also show that contemporaneous depth
adjustment can absorb persistence generated through replenishment.
Within the simulated mechanisms, the lag profile provides a
diagnostic of how persistence arises.

The application has several limitations. Five stocks observed
over one year provide limited evidence on cross-stock
heterogeneity. Although the cross-sectional pattern is consistent
with a role for relative tick size, the within-stock results do
not support it as a sufficient explanation. A broader sample and
exogenous tick-size changes would help assess this relationship.
The level comparisons are descriptive rather than causal, and
conditioning on visible depth does not identify the mechanism
behind any remaining difference. Entry depth also becomes less
informative as a visit lengthens; allowing its coefficient to
vary with elapsed time mitigates this issue in the simulations.
Finally, the findings concern the Nasdaq book and need not extend
to the consolidated market.

A further theoretical question concerns the joint convergence
of $(X^{(\delta)},\ell^{(\delta)})$ to a price process and its
local time as the tick size decreases, including the rate of
convergence. Extending the heuristic in \OAremlimit{} to allow
singular limits would help characterise sticky, skew and
oscillating behaviour through queue dynamics and establish
a theoretical basis for the lag-profile diagnostic.
 
\FloatBarrier


\begin{thebibliography}{99}
\footnotesize

\bibitem[Andersen and Gill(1982)]{AndersenGill1982}
Andersen, P.~K., and R.~D. Gill. 1982.
\newblock Cox's regression model for counting processes:
  A large sample study.
\newblock \emph{The Annals of Statistics} 10(4): 1100--1120.

\bibitem[Bai and Guo(2019)]{SkewIndex}
Bai, Y., and Z.~Guo. 2019.
\newblock An empirical investigation to the ``skew'' phenomenon
  in stock index markets: Evidence from the {N}ikkei 225 and others.
\newblock \emph{Sustainability} 11(24): 7219.

\bibitem[Borodin(1982)]{Borodin1981}
Borodin, A.~N. 1982.
\newblock An asymptotic behaviour of local times of a recurrent
  random walk with finite variance.
\newblock \emph{Theory of Probability \& Its Applications}
  26(4): 758--772.

\bibitem[Cont and de Larrard(2013)]{ContdeLarrard}
Cont, R., and A.~de Larrard. 2013.
\newblock Price dynamics in a Markovian limit order market.
\newblock \emph{SIAM Journal on Financial Mathematics}
  4(1): 1--25.

\bibitem[Delbaen and Shirakawa(2002)]{DelbaenShirakawa}
Delbaen, F., and H.~Shirakawa. 2002.
\newblock No arbitrage condition for positive diffusion price processes.
\newblock \emph{Asia-Pacific Financial Markets} 9: 159--168.

\bibitem[Engelbert and Peskir(2014)]{EngelbertPeskir}
Engelbert, H.-J., and G.~Peskir. 2014.
\newblock Stochastic differential equations for sticky Brownian motion.
\newblock \emph{Stochastics} 86(6): 993--1021.

\bibitem[Florens-Zmirou(1993)]{FlorensZmirou}
Florens-Zmirou, D. 1993.
\newblock On estimating the diffusion coefficient from discrete observations.
\newblock \emph{Journal of Applied Probability} 30(4): 790--804.

\bibitem[Harris(1991)]{Harris}
Harris, L. 1991.
\newblock Stock price clustering and discreteness.
\newblock \emph{Review of Financial Studies} 4(3): 389--415.

\bibitem[Huang et~al.(2015)]{HuangLehalleRosenbaum}
Huang, W., C.-A.~Lehalle, and M.~Rosenbaum. 2015.
\newblock Simulating and analyzing order book data:
  The queue-reactive model.
\newblock \emph{Journal of the American Statistical Association}
  110(509): 107--122.

\bibitem[Jacod(1998)]{JacodRates}
Jacod, J. 1998.
\newblock Rates of convergence to the local time of a diffusion.
\newblock \emph{Annales de l'I.H.P. Probabilit\'es et Statistiques}
  34(4): 505--544.

\bibitem[Jarrow et~al.(2011)]{JarrowKchiaProtter}
Jarrow, R., Y.~Kchia, and P.~Protter. 2011.
\newblock How to detect an asset bubble.
\newblock \emph{SIAM Journal on Financial Mathematics}
  2(1): 839--865.

\bibitem[Kalbfleisch and Prentice(2002)]{KalbfleischPrentice}
Kalbfleisch, J.~D., and R.~L.~Prentice. 2002.
\newblock \emph{The Statistical Analysis of Failure Time Data}.
\newblock 2nd ed. Wiley.

\bibitem[Knight(1963)]{Knight1963}
Knight, F.~B. 1963.
\newblock Random walks and a sojourn density process of {B}rownian motion.
\newblock \emph{Transactions of the American Mathematical Society}
  109(1): 56--86.

\bibitem[Lallouache and Abergel(2014)]{LallouacheAbergel}
Lallouache, M., and F.~Abergel. 2014.
\newblock Tick size reduction and price clustering in a FX order book.
\newblock \emph{Physica A} 416: 488--498.

\bibitem[Lejay(2006)]{Lejay}
Lejay, A. 2006.
\newblock On the constructions of the skew Brownian motion.
\newblock \emph{Probability Surveys} 3: 413--466.

\bibitem[Lejay et~al.(2014)]{LejayMordeckiTorres}
Lejay, A., E.~Mordecki, and S.~Torres. 2014.
\newblock Is a Brownian motion skew?
\newblock \emph{Scandinavian Journal of Statistics}
  41(2): 346--364.

\bibitem[Lin and Wei(1989)]{LinWei1989}
Lin, D.~Y., and L.~J. Wei. 1989.
\newblock The robust inference for the Cox proportional hazards model.
\newblock \emph{Journal of the American Statistical Association}
  84(408): 1074--1078.

\bibitem[Maeda and Jacka(2017)]{OptimalStopping}
Maeda, J., and S.~D.~Jacka. 2017.
\newblock An optimal stopping problem modeling technical analysis.
\newblock arXiv:1707.05253. Revised 2020.

\bibitem[Mazzonetto(2026)]{MazzonettoLejay}
Mazzonetto, S. 2026.
\newblock Rates of convergence to the local time of oscillating
  and skew {B}rownian motions.
\newblock \emph{Bernoulli} 32(1): 567--589.

\bibitem[Osborne(1962)]{Osborne}
Osborne, M.~F.~M. 1962.
\newblock Periodic structure in the Brownian motion of stock prices.
\newblock \emph{Operations Research} 10(3): 345--379.

\bibitem[Revuz and Yor(1999)]{RevuzYor}
Revuz, D., and M.~Yor. 1999.
\newblock \emph{Continuous Martingales and Brownian Motion}.
\newblock 3rd ed. Springer.

\bibitem[Robert and Rosenbaum(2011)]{RobertRosenbaum}
Robert, C.~Y., and M.~Rosenbaum. 2011.
\newblock A new approach for the dynamics of ultra-high-frequency data:
  The model with uncertainty zones.
\newblock \emph{Journal of Financial Econometrics}
  9(2): 344--366.

\end{thebibliography}
\end{document}